\documentclass{article}
\usepackage{amsfonts}
\usepackage{amsmath}

\newtheorem{theorem}{Theorem}

\newtheorem{remark}[theorem]{Remark}

\input{tcilatex}
\begin{document}

\title{Generalized Labeled Multi-Bernoulli Filters\\
and Multitarget-Correlation Models}
\author{Ronald Mahler, Eagan MN, U.S.A., mahlerronald@comcast.net}
\maketitle

\begin{abstract}
The generalized labeled multi-Bernoulli (GLMB)\ filter is a theoretically
rigorous Bayes-optimal multitarget tracking algorithm with computationally
tractable implementations, based on labeled random finite set (LRFS)\
theory. \ It presumes that multitarget populations can be approximated using
GLMB multitarget probability density functions (p.d.f.'s), which consist of
weighted hypotheses regarding the current target-states. \ A special case of
the GLMB\ p.d.f.---the LMB p.d.f.---presumes that the targets are
statistically independent. \ This paper demonstrates that a) GLMB p.d.f.'s
can be interpreted as straightforward generalizations of LMB p.d.f.'s to
statistically correlated target populations, given an implicit presumption
of \textquotedblleft simple labeled correlation\textquotedblright\ (SLC)
models of multitarget correlation; b) the GLMB\ filter can be reformulated
as a SLC-GLMB filter; and c) SLC\ models seem primarily appropriate for
target-clusters consisting of small numbers of closely-spaced targets.
\end{abstract}

\section{Introduction \label{A-Intro}}

The generalized labeled multi-Bernoulli (GLMB)\ filter is a theoretically
rigorous Bayes-optimal multitarget tracking algorithm with computationally
tractable implementations, all based on labeled random finite set (LRFS)\
theory. \ Relevant publications, listed chronologically, include \cite%
{Vo-ISSNIP12-Conjugate}, \cite{VoVoSPIE13}, \cite{VoPhungTSP2014}, \cite%
{VoVoTSPEfficient}, \cite[Chpt. 15]{Mah-Newbook} \cite{VoTSPMultiscan2019}, 
\cite{Beard2020}, \cite{ShimVo-TSP2023}, and especially, the overview paper 
\cite{Vo-TSP-Overview2024}. \ 

The GLMB\ filter presumes that multitarget populations can be approximated
using GLMB multitarget probability density functions (p.d.f.'s), which
incorporate weighted hypotheses regarding the current target-states. \ A
special case of the GLMB\ p.d.f.---the LMB p.d.f.---presumes that the
targets are statistically independent.

The main results of this paper are the following:

\begin{enumerate}
\item GLMB p.d.f.'s can be interpreted as straightforward generalizations of
LMB p.d.f.'s to statistically correlated target populations, based on an
implicit presumption of \textquotedblleft simple labeled
correlation\textquotedblright\ (SLC) models of multitarget correlation
(Remark \ref{Rem-GenLMB}). \ 

\item SLC\ models:

\begin{enumerate}
\item represent the kinematic uncertainty of individual targets using
random---rather than the usual deterministic---spatial p.d.f.'s (Eq. (\ref%
{eq-Random}));

\item generalize GLMB multitarget state models to account for \textit{a
priori} multitarget correlation---i.e., correlations not attributable to
collected measurements; and thus

\item appear to be most appropriate for addressing target-clusters
consisting of small numbers of closely-spaced targets (Section \ref%
{A-Application}).\footnote{%
If a cluster has at most two targets then it can be optimally addressed
using the \textquotedblleft dyadic filter\textquotedblright\ of \cite%
{Mahler-IETDyad-2022}. \ }
\end{enumerate}

\item The GLMB\ filter can be reformulated as a SLC-GLMB filter (Section \ref%
{A-SLCGLMB}).
\end{enumerate}

Begin with the simplest instance of an SLC\ model: \ correlation between two
targets whose respective labels are \ $l_{1},l_{2}$ \ with \ $l_{1}\neq
l_{2} $: \ 
\begin{equation}
\mathring{s}_{L}(x_{1},x_{2})=\sum_{i\in \mathbb{I}}\alpha _{i}^{L}\,%
\mathring{s}_{l_{1}}^{i}(x_{1})\,\mathring{s}_{l_{2}}^{i}(x_{2})
\end{equation}%
where\ a) \ $L=\{l_{1},l_{2}\}$; \ b) $\ \mathbb{I}$ \ is a finite,
non-empty index-set; c) \ $\alpha _{i}^{L}\geq 0$ \ and \ $\sum_{i\in 
\mathbb{I}}\alpha _{i}^{L}=1$;\ and d) \ $\mathring{s}_{l_{1}}^{i}(x)$, $\ \,%
\mathring{s}_{l_{2}}^{i}(x)$ \ for \ $i\in \mathbb{I}$ \ are hypothesized
single-target spatial p.d.f.'s for targets \ $l_{1}$ \ and \ $l_{2}$, \
respectively. \ 

\begin{remark}
\label{Rem-Context}In the context of GLMB filters, the \ $L$ \ are finite,
nonempty label-sets and the \ $\mathbb{I}$ \ are nonempty finite sets of
time-sequences \ $\mathring{\theta}_{1:k}:\mathring{\theta}_{1},...,%
\mathring{\theta}_{k}$, where $\ \mathring{\theta}_{1},...,\mathring{\theta}%
_{k}$\ \ are measurement-to-track associations (MTAs)\ at times \ $%
t_{1},...,t_{k}$---see Section \ref{a-SLC-AA-Example}.
\end{remark}

The marginal p.d.f.'s of \ $\mathring{s}_{L}(x_{1},x_{2})$ \ are the
averaged spatial p.d.f.'s \ 
\begin{equation}
\mathring{s}_{l_{1}|L}(x_{1})=\sum_{i\in \mathbb{I}}\alpha _{i}^{L}\,%
\mathring{s}_{l_{1}}^{i}(x_{1})\text{, \ \ \ }\mathring{s}%
_{l_{2}|L}(x_{2})=\sum_{i\in \mathbb{I}}\alpha _{i}^{L}\,\mathring{s}%
_{l_{2}}^{i}(x_{2}).
\end{equation}

The statistical correlation between two LRFSs can be measured using the 
\textit{factorial covariance density} (f.c.d., see Section \ref%
{A-Math-AA-FacCov}). \ In the present case the f.c.d. reduces to%
\begin{equation}
\mathring{c}_{L}^{[2]}(\mathring{x}_{1}^{\prime },\mathring{x}_{2}^{\prime
})=\delta _{L,L^{\prime }}\left( \mathring{s}_{L^{\prime }}(x_{1}^{\prime
},x_{2}^{\prime })-\mathring{s}_{l_{1}^{\prime }|L^{\prime }}(x_{1}^{\prime
})\,\mathring{s}_{l_{2}^{\prime }|L^{\prime }}(x_{2}^{\prime })\right)
\end{equation}%
(see Eq. (\ref{eq-FacCov-2})) where \ a)\ $\mathring{x}_{1}^{\prime
}=(x_{1}^{\prime },l_{1}^{\prime })$, \ $\mathring{x}_{2}^{\prime
}=(x_{2}^{\prime },l_{2}^{\prime })$: b) \ $l_{1}^{\prime }\neq
l_{2}^{\prime }$\ and \ $L^{\prime }=\{l_{1}^{\prime },l_{2}^{\prime }\}$;\
and c) \ $\delta _{L,L^{\prime }}$ \ is a Kronecker delta\footnote{%
That is, \ $\delta _{L,L^{\prime }}=1$ \ if \ $L=L^{\prime }$ \ and \ $%
\delta _{L,L^{\prime }}=0$ \ otherwise.}.\ 

Suppose that the random vectors \ $\mathbf{x}_{l_{1}},\mathbf{x}_{l_{2}}$
have respective p.d.f.'s \ $\mathring{s}_{l_{1}|L}$ \ and \ $\mathring{s}%
_{l_{2}|L}$ \ and joint distribution \ $\mathring{s}_{L}$. \ Then \ $\mathbf{%
x}_{l_{1}},\mathbf{x}_{l_{2}}$ \ are statistically independent if and only
if the f.c.d. vanishes identically. \ Because of this (and also because SLC\
models are a central aspect of the GLMB filter), it follows that the SLC
approach offers theoretically reasonable modeling of multitarget correlation.

The remainder of the paper is organized as follows. \ Mathematical
background (Section \ref{A-Math}); enhanced GLMB RFSs (Section \ref{A-Aug});
SLC models (Section \ref{A-SLC}); the SLC-GLMB\ filter (Section \ref%
{A-SLCGLMB}); a proposed application to closely-spaced targets (Section \ref%
{A-Application}); mathematical derivations (Section \ref{A-Der}); and
Conclusions (Section \ref{A-Concl}).

\section{Mathematical Background \label{A-Math}}

The section is organized as follows: \ Basic notation (Section \ref%
{A-Math-AA-Nota}); LRFSs (Section \ref{A-Math-AA-LRFS}); set integrals
(Section \ref{A-Math-AA-SetInt}); the multitarget recursive Bayes filter
(Section \ref{A-Math-AA-MRBF}); probability generating functionals (Section %
\ref{A-Math-AA-PGFL}); functional derivatives (Section \ref%
{A-Math-AA-FuncDer}); GLMB RFSs (Section \ref{A-Math-AA-GLMBDistr}); LMB\
RFSs (Section \ref{A-Math-AA-LMB}); the GLMB\ filter (Section \ref%
{A-Math-AA-GLMB}); and factorial covariance densities (Section \ref%
{A-Math-AA-FacCov}).

\subsection{Basic Notation \label{A-Math-AA-Nota}}

In what follows, let us be given:

\begin{enumerate}
\item A single-target measurement space \ $\mathbb{Z}$ \ with measurements \ 
$z\in \mathbb{Z}$.\ 

\item A single-target state space \ $\mathbb{\mathring{X}}=\mathbb{X}%
_{0}\times \mathbb{L}$ \ with elements \ $\mathring{x}=(x,l)\in \mathbb{%
\mathring{X}}$, \ where \ $\mathbb{X}_{0}$ \ is the kinematic state space
and \ $\mathbb{L}$ \ is some finite subset of a countably infinite set of
\textquotedblleft labels\textquotedblright\ $\ l$. \ Labels are provisional
stand-ins for a discrete target state variable: \ unique target identity.

\item An integral on \ $\mathbb{\mathring{X}}$, \ 
\begin{equation}
\int \mathring{f}(\mathring{x})d\mathring{x}\overset{_{\text{def.}}}{=}%
\sum_{l\in \mathbb{L}}\int \mathring{f}(x,l)dx.\ 
\end{equation}%
If \ $\mathring{f}(\mathring{x})\geq 0$ \ and \ $\int \mathring{f}(\mathring{%
x})d\mathring{x}=1$ then \ $\mathring{f}(\mathring{x})$ \ is a p.d.f. on \ $%
\mathbb{\mathring{X}}$. \ In this case, the integral is mathematically
well-defined if and only if the unit of measurement (UoM)\ of \ $\mathring{f}%
(\mathring{x})$\ \ is \ $\iota _{\mathbb{X}_{0}}^{-1}$ \ where \ $\iota _{%
\mathbb{X}_{0}}$ \ is the UoM of \ $\mathbb{X}_{0}$.

\item At time \ $t_{k}$, a single-target measurement density (a.k.a.
likelihood function) $f_{k}(z|\mathring{x})=L_{z}(\mathring{x})$, and a
single-target Markov density $\ $%
\begin{equation}
M_{(x,l)}(x^{\prime },l^{\prime })=f_{k|k-1}(x,l|x^{\prime },l^{\prime
})=\delta _{l,l^{\prime }}\,f_{k|k-1}(x|x^{\prime },l^{\prime })=\delta
_{l,l^{\prime }}\,M_{x}(x^{\prime },l^{\prime }).\ 
\end{equation}%
The last of these equalities ensures that target labels are preserved during
a Markov state-transition. \ 

\item At time \ $t_{k}$, \ a single-target probability of detection \ $%
\mathring{p}_{D}(\mathring{x})\overset{_{\text{abbr.}}}{=}\mathring{p}_{D,k}(%
\mathring{x})$; a probability of target survival \ $\mathring{p}_{S}(%
\mathring{x})\overset{_{\text{abbr.}}}{=}\mathring{p}_{S,k|k-1}(\mathring{x}%
) $; and an intensity function \ $\kappa _{k}(z)$ \ for a Poisson clutter
process. \ \ 
\end{enumerate}

\subsection{Labeled Random Finite Sets \label{A-Math-AA-LRFS}}

Let \ $\mathring{X}\subseteq \mathbb{\mathring{X}}$ \ be a finite set, let \ 
$|\mathring{X}|$ \ denote the number of elements in \ $\mathring{X}$, and,
if \ $\mathring{X}=\{(x_{1},l_{1}),...,(x_{n},l_{n})\}$, \ let \ $\mathring{X%
}_{\mathbb{L}}=\{l_{1},...,l_{n}\}$ \ denote the set of labels in \ $%
\mathring{X}$.

Given this, \ $\mathring{X}$ \ is a \textit{labeled finite set }(LFS) if \ $|%
\mathring{X}_{\mathbb{L}}|=|\mathring{X}|\,$---i.e., if the elements of \ $%
\mathring{X}$ \ have distinct labels. \ If \ $\mathring{X}$ \ is not an LFS
then it is physically impossible---e.g., $\ \mathring{X}%
=\{(x_{1},l),(x_{2},l)\}$ \ with \ $x_{1}\neq x_{2}$.\footnote{%
Since point targets have no physical extent, \ $\{(x,l_{1}),(x,l_{2})\}$
with \ $l_{1}\neq l_{2}$ \ is a valid LFS.} \ The mathematical states of a
multitarget system are the LFSs.

A \textit{labeled random finite set} (LRFS) \ $\mathring{\Xi}$\ \ is an
random variable whose realizations are LFSs. \ It is characterized by its
probability p.d.f. \ $\mathring{f}_{\mathring{\Xi}}(\mathring{X})\geq 0$,
which has the property that \ $\mathring{f}_{\mathring{\Xi}}(\mathring{X})=0$
\ if \ $|\mathring{X}_{\mathbb{L}}|\neq |\mathring{X}|\,$---i.e., if \ $%
\mathring{X}$ \ is physically impossible. \ 

In more detail:\ \ If \ $l_{1},...,l_{n}$ \ are not distinct then \ $%
\mathring{f}_{\mathring{\Xi}}(\{(x_{1},l_{1}),...,(x_{n},l_{n})\})=0$. \ The
unit of measurement (UoM) of \ $\mathring{f}_{\mathring{\Xi}}(\mathring{X})$
\ is \ $\iota _{\mathbb{X}_{0}}^{-|\mathring{X}|}$ \ if \ $\iota _{\mathbb{X}%
_{0}}$ \ is the UoM of \ $\mathbb{X}_{0}$. \ 

\subsubsection{Vo-Vo Labels \label{A-Math-AA-VoVo}}

Denote the set of positive integers as \ $\mathbb{N}_{+}=\{1,2,...\}$ \ with
\ $i\in \mathbb{N}$ \ and the set of nonnegative integers as \ $\mathbb{N}%
=\{0,1,2,...\}$ \ with $k\in \mathbb{N}$. \ Then \cite[Eq. (15.148,15.149)]%
{Mah-Newbook}: a) \ $\mathbb{L}=\mathbb{N}\times \mathbb{N}_{+}$ \ is the
set of possible Vo-Vo target labels; b) \ \ 
\begin{equation}
\mathbb{L}_{0:k}=\{0,1,...,k-1\}\times \mathbb{N}  \label{eq-Label-Prev}
\end{equation}%
(with Cartesian product \ \textquotedblleft $\times $\textquotedblright ) is
the subset of possible labels \ $(j,i)$ \ for the targets that appeared at
times $\ t_{j}$ \ for \ $j=0,1,...,k-1$; and\ c) \ \ 
\begin{equation}
\mathbb{L}_{k}=\{k\}\times \mathbb{N}  \label{eq-Label-New}
\end{equation}%
is the set of possible labels for the targets that appear at time \ $t_{k}$.

\subsection{Set Integrals \label{A-Math-AA-SetInt}}

Let \ $\mathring{f}(\mathring{X})\geq 0$ \ be any function of an LFS
variable \ $\mathring{X}\subseteq \mathbb{\mathring{X}}$ \ such that the UoM
of \ $\mathring{f}(\mathring{X})$ \ is \ $\iota _{\mathbb{X}_{0}}^{-|%
\mathring{X}|}$. \ Then its \textit{set integral} is%
\begin{equation}
\int \mathring{f}(\mathring{X})\delta \mathring{X}=\sum_{n\geq 0}\frac{1}{n!}%
\sum_{(l_{1},...,l_{n})\in \mathbb{L}^{n}}\int
f(\{(x_{1},l_{1}),...,(x_{n},l_{n})\})dx_{1}\cdots dx_{n}
\label{eq-SetIntLabeled}
\end{equation}%
where, by convention, \ $\mathring{f}(\{(x_{1},l_{1}),...,(x_{n},l_{n})\})=%
\mathring{f}(\emptyset )$ \ if \ $n=0$. \ Thus \ $\mathring{f}(\mathring{X})$
\ is a multitarget p.d.f. if and only if \ $\int \mathring{f}(\mathring{X}%
)\delta \mathring{X}=1$.

\subsection{The Multitarget Recursive Bayes Filter \label{A-Math-AA-MRBF}}

The (labeled) multitarget recursive Bayes (MTRB) filter is given by 
\begin{eqnarray}
\mathring{f}_{k|k-1}(\mathring{X}|Z_{1:k-1}) &=&\int \mathring{f}_{k|k-1}(%
\mathring{X}|\mathring{X}_{k-1})\,f_{k-1|k-1}(\mathring{X}%
_{k-1}|Z_{1:k-1})\delta \mathring{X}_{k-1}  \label{eq-MTRBF-P} \\
\mathring{f}_{k|k}(\mathring{X}|Z_{1;k}) &=&\frac{f_{k}(Z_{k}|\mathring{X}%
)\,f_{k|k-1}(\mathring{X}|Z_{1:k-1})}{f_{k}(Z_{k}|Z_{1:k-1})}
\label{eq-MTRBF-B} \\
f_{k}(Z_{k}|Z_{1:k-1}) &=&\int f_{k}(Z_{k}|\mathring{X})\,f_{k|k-1}(%
\mathring{X}|Z_{1:k-1})\delta \mathring{X}.  \label{eq-MTRBF-normal}
\end{eqnarray}%
with initial distribution \ $f_{0|0}(\mathring{X})$. \ Here,\ a) \ $f_{k|k}(%
\mathring{X}|Z_{1;k})$ \ is the probability (density) that the finite set \ $%
\mathring{X}\subseteq \mathbb{\mathring{X}}$ \ of targets is present at time
\ $t_{k}$ \ if the sequence \ $Z_{1;k}:Z_{1},...,Z_{k}$ \ of
measurement-sets has been collected; b) $f_{k}(Z|\mathring{X})$ \ is the
probability (density) that the measurement-set \ $Z$ \ will collected at
time \ $t_{k}$ \ if \ $\mathring{X}$ \ is present at time \ $t_{k}$; c) $\ 
\mathring{f}_{k|k-1}(\mathring{X}|\mathring{X}_{k-1})$ \ is the probability
(density) that the state-set \ $\mathring{X}_{k-1}$ \ at time \ $t_{k-1}$ \
will evolve to the state-set \ $\mathring{X}$ \ at time \ $t_{k}$; d) $\
f_{k}(Z|\mathring{X})=L_{Z}(\mathring{X})$ \ is the multitarget likelihood
function; and e) \ $f_{k}(Z|\emptyset )=\kappa _{k}(Z)$ \ where \ $\kappa
_{k}(Z)$ \ is the multi-object p.d.f. of the clutter process.

\subsection{Probability Generating Functionals \label{A-Math-AA-PGFL}}

The probability generating functional (p.g.fl.) of an LRFS \ $\mathring{\Xi}$
\ is%
\begin{equation}
\mathring{G}_{\mathring{\Xi}}[\mathring{h}]=\int \mathring{h}^{\mathring{X}%
}\,\mathring{f}_{\mathring{\Xi}}(\mathring{X})\delta \mathring{X}
\end{equation}%
where, for any unitless \textquotedblleft test function\textquotedblright\ \ 
$0\leq \mathring{h}(\mathring{x})\leq 1$, \ $\mathring{h}^{\mathring{X}}=1$
\ if \ $\mathring{X}=\emptyset $ \ and\ \ $\mathring{h}^{\mathring{X}%
}=\prod_{\mathring{x}\in \mathring{X}}\mathring{h}(\mathring{x})$ \ if
otherwise. \ Thus \ $\mathring{G}_{\mathring{\Xi}}[\mathring{h}]$ \ is
unitless and \ $0\leq \mathring{G}_{\mathring{\Xi}}[\mathring{h}]\leq 1$.

Let \ $\mathring{s}(\mathring{x})=\mathring{s}(x,l)\geq 0$ \ be a density
function of the variable \ $\mathring{x}=(x,l)\in \mathbb{\mathring{X}}$%
---i.e., its UoM is $\ \iota _{\mathbb{X}_{0}}^{-1}$. \ In what follows it
will be assumed that \ $\int \mathring{s}(x,l)dx=1$ \ for all \ $l\in 
\mathbb{L}$---i.e., \ $\mathring{s}_{l}(x)\overset{_{\text{def.}}}{=}%
\mathring{s}(x,l)$\ is the spatial p.d.f. of the target \ $l$. \ In this
case the following functional notation will frequently be used: 
\begin{equation}
\mathring{s}_{l}[\mathring{h}]\overset{_{\text{abbr.}}}{=}\int \mathring{h}%
(x,l)\,\mathring{s}(x,l)dx.  \label{eq-FuncNotation}
\end{equation}%
\ Note that \ $\mathring{s}_{l}[\mathring{h}]$ \ is unitless and linear: 
\begin{equation}
\ \mathring{s}_{l}[a_{1}\mathring{h}_{1}+a_{2}\mathring{h}_{2}]=a_{1}\,%
\mathring{s}_{l}[h_{1}]+a_{2}\,\mathring{s}_{l}[\mathring{h}_{2}].\ 
\end{equation}

\subsection{Functional Derivatives \label{A-Math-AA-FuncDer}}

Let \ $\mathring{F}[\mathring{h}]\geq 0$ \ be a functional of a
test-function variable $\mathring{h}$\ and \ $\mathring{X}$ \ an LFS. \ Then
the functional derivative of \ $\mathring{F}$ \ with respect to \ $\mathring{%
X}$ \ is%
\begin{equation}
\frac{\delta \mathring{F}}{\delta \mathring{X}}[\mathring{h}]=\left\{ 
\begin{array}{ccc}
\mathring{F}[\mathring{h}] & \text{if} & \mathring{X}=\emptyset \\ 
\lim_{\varepsilon \rightarrow 0^{+}}\frac{\mathring{F}[\mathring{h}%
+\varepsilon \delta _{\mathring{x}}]-\mathring{F}[\mathring{h}]}{\varepsilon 
} & \text{if} & \mathring{X}=\{\mathring{x}\} \\ 
\frac{\delta ^{n}\mathring{F}}{\delta \mathring{x}_{1}\cdots \delta 
\mathring{x}_{1}}[\mathring{h}] & \text{if} & |\mathring{X}|=n\geq 2%
\end{array}%
\right.
\end{equation}%
where \ $\delta _{(x,l)}(x^{\prime },l^{\prime })=\delta _{l,l^{\prime
}}\,\delta _{x}(x^{\prime })$ \ is the Dirac delta function concentrated at
\ $(x,l)$.\footnote{%
The limit definition in the \ $\mathring{X}=\{\dot{x}\}$ \ case is
heuristic. \ See \cite{Mah-Artech} for a rigorous definition.}

It can be shown that the p.g.fl. and p.d.f. of \ $\mathring{\Xi}$ \ are
related by \ 
\begin{equation}
\mathring{f}_{\mathring{\Xi}}(\mathring{X})=\frac{\delta \mathring{G}_{%
\mathring{\Xi}}}{\delta \mathring{X}}[0]=\left[ \frac{\delta \mathring{G}_{%
\mathring{\Xi}}}{\delta \mathring{X}}[\mathring{h}]\right] _{\mathring{h}=0}%
\text{.}
\end{equation}

As a simple example,%
\begin{equation}
\frac{\delta }{\delta (x^{\prime },l^{\prime })}\mathring{s}_{l}[\mathring{h}%
]=\delta _{l,l^{\prime }}\,\mathring{s}_{l^{\prime }}(x^{\prime }).
\label{eq-Example}
\end{equation}

\subsection{GLMB\ RFSs \label{A-Math-AA-GLMBDistr}}

The p.d.f. of a GLMB\ RFS has the form \cite[Eq. (15.112)]{Mah-Newbook} \ 
\begin{equation}
\mathring{f}_{\mathring{\Xi}}(\mathring{X})=\delta _{|\mathring{X}|,|%
\mathring{X}_{\mathbb{L}}|}\sum_{o\in \mathbb{O}}\omega ^{o}(\mathring{X}_{%
\mathbb{L}})\,(\mathring{s}^{o})^{\mathring{X}}  \label{eq-GLMB-Distr}
\end{equation}%
where: \ a) $\mathbb{O}$ \ is a finite, nonempty index set; b) \ $0\leq
\omega ^{o}(L)\leq 1$; \ c) \ $\omega ^{o}(L)>0$ \ for only a finite number
of pairs \ $(o,L)$ \ such that \ $o\in \mathbb{O}$ \ and $L\subseteq \mathbb{%
L}$; d) \ $\sum_{o\in \mathbb{O}}\sum_{L\subseteq \mathbb{L}}\omega
^{o}(L)=1 $; and e) $\mathring{s}_{l}^{o}(x)=\mathring{s}^{o}(x,l)$ \ is the
spatial distribution of the target \ $l\in \mathring{X}_{\mathbb{L}}$%
---i.e., $\int \mathring{s}_{l}^{o}(x)dx=1$.\ 

The p.g.fl. corresponding to Eq. (\ref{eq-GLMB-Distr}) is \cite[Eq. (15.115)]%
{Mah-Newbook} 
\begin{equation}
\mathring{G}_{\mathring{\Xi}}[\mathring{h}]=\sum_{o\in \mathbb{O}%
}\sum_{L\subseteq \mathbb{L}}\omega ^{o}(L)\prod_{l\in L}\mathring{s}%
_{l}^{o}[\mathring{h}]  \label{eq-GLMB-PGFL}
\end{equation}%
where \ $\mathring{s}_{l}^{o}[\mathring{h}]$ is as in Eq. (\ref%
{eq-FuncNotation}).\ 

Note that Eq. (\ref{eq-GLMB-Distr}) can be alternatively written as%
\begin{eqnarray}
\mathring{f}_{\mathring{\Xi}}(\mathring{X}) &=&\delta _{|\mathring{X}|,|%
\mathring{X}_{\mathbb{L}}|}\sum_{o\in \mathbb{O}}\omega ^{o}(\mathring{X}_{%
\mathbb{L}})\prod_{l\in \mathring{X}_{\mathbb{L}}}\,\mathring{s}^{o}(%
\mathring{X}_{\ast l})  \label{eq-GLMB-Distr-Alt1} \\
&=&\delta _{|\mathring{X}|,|\mathring{X}_{\mathbb{L}}|}\sum_{o\in \mathbb{O}%
}\omega ^{o}(\mathring{X}_{\mathbb{L}})\prod_{l\in \mathring{X}_{\mathbb{L}%
}}\,\mathring{s}_{l}^{o}(\mathring{X}_{l})  \label{eq-GLMB-Distr-Alt2}
\end{eqnarray}%
where, for all \ $l\in \mathring{X}_{\mathbb{L}}$, the elements \ $\mathring{%
X}_{\ast l}\in \mathbb{X}_{0}\times \mathbb{L}$ \ and\ \ $\mathring{X}%
_{l}\in \mathbb{X}_{0}$ \ are implicitly and uniquely defined in terms of
the LFS \ $\mathring{X}$\ \ by the relations \ 
\begin{equation}
\mathring{X}_{\ast l}=(\mathring{X}_{l},l)\in \mathring{X}\cap (\mathbb{X}%
_{0}\times \{l\}).  \label{eq-Big}
\end{equation}

\subsection{LMB\ RFSs \label{A-Math-AA-LMB}}

LMB RFSs are a special case of GLMB RFSs. \ They model populations of
statistically independent targets that have explicit probabilities of
existence.

Let \ $J\subseteq \mathbb{L}$ \ and let\ \ $0\leq q_{l}\leq 1$ \ denote the
probability of existence of the target with label \ $l\,\in J$. \ Then the\
p.d.f. of the corresponding LMB\ RFS has the form \cite[Eq. (15.87)]%
{Mah-Newbook} 
\begin{equation}
\mathring{f}_{J}(\mathring{X})=\delta _{|\mathring{X}|,|\mathring{X}_{%
\mathbb{L}}|}\,\omega _{J}(\mathring{X}_{\mathbb{L}})\,\mathring{s}^{%
\mathring{X}}  \label{eq-LMB-1}
\end{equation}%
where, for \ $L\subseteq \mathbb{L}$, \ \ 
\begin{equation}
\omega _{J}(L)=\left( \prod_{l\in J-L}(1-q_{l})\right) \left( \prod_{l\in
L}q_{l}\mathbf{1}_{J}(l)\right)  \label{eq-LMB-2}
\end{equation}%
where \ $\mathbf{1}_{J}(l)$ \ is the set-indicator function of \ $J$ \ and
where \ $\mathring{s}_{l}(x)=\mathring{s}(x,l)$ \ with \ $l\in J$ \ are the
corresponding spatial distributions. \ Note that because of the factor \ $%
\mathbf{1}_{J}(l)$, \ $\omega _{J}(L)=0$ \ if \ $L$ \ is not a subset of $J$%
. \ 

\begin{remark}
\label{Rem-Special}If all targets in \ $J$\ \ exist---i.e., \ $q_{l}=1$ \
for all \ $l\in J$---then \ $\omega _{J}(L)$ \ reduces to \ $\omega
_{J}(L)=\delta _{J,\mathring{X}_{\mathbb{L}}}$ and so $\mathring{f}_{J}(%
\mathring{X})$ \ reduces to \ 
\begin{equation}
\mathring{f}_{J}(\mathring{X})=\delta _{|\mathring{X}|,|\mathring{X}_{%
\mathbb{L}}|}\,\delta _{J,\mathring{X}_{\mathbb{L}}}\mathring{s}^{\mathring{X%
}}=\delta _{|\mathring{X}|,|\mathring{X}_{\mathbb{L}}|}\,\delta _{J,%
\mathring{X}_{\mathbb{L}}}\,\mathring{s}^{\mathring{X}}.
\label{eq-SpecialLMB}
\end{equation}
\end{remark}

The p.g.fl. corresponding to Eq. (\ref{eq-LMB-1}) is \ 
\begin{equation}
\mathring{G}_{J}[\mathring{h}]=\prod_{l\in J}\left( 1-q_{l}+q_{l}\,s_{l}[%
\mathring{h}]\right) .
\end{equation}

An LMB\ RFS has the form \ $\mathring{\Xi}=\mathring{\Xi}_{1}\cup ...\cup 
\mathring{\Xi}_{n}$ \ where the \ $\mathring{\Xi}_{i}$ \ are mutually
independent Bernoulli LRFSs---i.e., \ LMB RFSs with \ $|J_{i}|=1$ \ for \ $%
i=1,...,n$. \ The p.d.f. and p.g.fl. of a Bernoulli LRFS for a target with
label \ $l$ \ are, respectively,%
\begin{eqnarray}
\mathring{f}_{\{l\}}(\mathring{X}) &=&\left\{ 
\begin{array}{ccc}
1-q_{l} & \text{if} & X=\emptyset \\ 
\delta _{l,l^{\prime }}\,q_{l^{\prime }}\,s_{l^{\prime }}(x^{\prime }) & 
\text{if} & X=\{(x^{\prime },l^{\prime })\} \\ 
0 & \text{if} & |X|\geq 2%
\end{array}%
\right.  \label{eq-Bern-Labeled} \\
\mathring{G}_{\{l\}}[\mathring{h}] &=&1-q_{l}+q_{l}\,s_{l}[\mathring{h}].
\end{eqnarray}

\subsection{The GLMB\ Filter \label{A-Math-AA-GLMB}}

The GLMB filter is an algebraically exact closed-form\footnote{%
In the sense of \cite{MahSensors1019Exact}.} special case (\textquotedblleft
solution\textquotedblright ) of the MRBF, assuming that \ $\mathring{f}%
_{k|k-1}(\mathring{X}|Z_{1:k-1})$ \ and \ $\mathring{f}_{k|k}(\mathring{X}%
|Z_{1:k})$ \ are, approximately, GLMB p.d.f.'s and \ $\mathring{f}_{0|0}(%
\mathring{X})$ \ is, approximately, an LMB\ p.d.f. \ 

The label-sets for \ $\mathring{f}_{k|k-1}(\mathring{X}|Z_{1:k-1})$ \ and \ $%
\mathring{f}_{k|k}(\mathring{X}|Z_{1:k})$ \ are, respectively, denoted as \ $%
L_{k|k-1}$ \ and $L_{k|k}$. \ The index sets for \ \ $\mathring{f}_{k|k-1}(%
\mathring{X}|Z_{1:k-1})$ \ and \ $\mathring{f}_{k|k}(\mathring{X}|Z_{1:k})$
\ are, respectively, denoted as \ $\mathbb{O}_{k|k-1}$ \ and \ $\mathbb{O}%
_{k|k}$ \ with \ $\mathbb{O}_{k|k-1}=\mathbb{O}_{k|k}$. \ As will be
presently explained, \ $\mathbb{O}_{k|k}=\mathcal{T}_{Z_{1}}\times ...\times 
\mathcal{T}_{Z_{k}}$\ \ where \ $\mathcal{T}_{Z_{i}}$ \ with \ $i=1,...,k$ \
is a finite set of measurement-to-track associations (MTAs) at time \ $t_{i}$%
.

The subsection is organized as follows: \ GLMB filter time-update (Section %
\ref{A-Math-AA-GLMB-AAA-P}); and GLMB filter measurement-update (Section \ref%
{A-Math-AA-GLMB-AAA-B}). \ 

\subsubsection{GLMB Filter Time-Update \label{A-Math-AA-GLMB-AAA-P}}

This subsection summarizes and clarifies the somewhat unclear discussion in 
\cite{Mah-Newbook}, pp. 483-484, 489-495.

The set of labels of the time-updated targets at time \ $t_{k}$ \ is \ 
\begin{equation}
L_{k|k-1}=L_{k|k-1}^{B}\uplus L_{k-1|k-1}\subseteq \mathbb{L}_{0:k}\ 
\end{equation}%
where \ $L_{k|k-1}^{B}\subseteq \mathbb{L}_{k}$ \ are the labels of the
\textquotedblleft birth\textquotedblright\ (newly-appearing) targets at time 
$t_{k}$; and where \ $L_{k-1|k-1}$ \ is the set of labels of the persisting
measurement-updated targets at time \ $t_{k-1}$.\footnote{%
In \cite{Mah-Newbook}, pp. 483-484, 489-495, the formula \textquotedblleft $%
L\subseteq \mathbb{L}_{0:k}$\textquotedblright\ \ was used instead of the
formula\ \textquotedblleft $L\subseteq L_{k|k-1}$\textquotedblright\ used in
what follows. \ While \ \textquotedblleft $L\subseteq \mathbb{L}_{0:k}$%
\textquotedblright\ \ is mathematically correct, \ \textquotedblleft $%
L\subseteq L_{k|k-1}$\textquotedblright\ is more specific.} \ 

The single-target labeled Markov density is \ 
\begin{equation}
\mathring{f}_{k|k-1}(x,l|x^{\prime }l^{\prime })=\delta _{l,l^{\prime }}\,%
\mathring{f}_{k|k-1}(x|x^{\prime }l^{\prime })=\delta _{l,l^{\prime }}\,%
\mathring{M}_{x}(x^{\prime },l^{\prime }).
\end{equation}%
That is, target labels are preserved under time-update.\ \ 

Let us be given the posterior GLMB distribution at time \ $t_{k-1}$ \cite[%
Eq. (15.229)]{Mah-Newbook},%
\begin{equation}
\mathring{f}_{k-1|k-1}(\mathring{X})=\delta _{|\mathring{X}|,|\mathring{X}_{%
\mathbb{L}}|}\,\sum_{o\in \mathbb{O}_{k-1|k-1}}\omega _{k-1|k-1}^{o}(%
\mathring{X}_{\mathbb{L}})\,(\mathring{s}_{k-1|k-1})^{\mathring{X}},
\end{equation}%
with label-set \ $L_{k-1|k-1}\subseteq \mathbb{L}_{0:k-1}$ and target
spatial distributions \ $\mathring{s}_{l^{\prime },k-1|k-1}(x^{\prime })=%
\mathring{s}_{k-1|k-1}(x^{\prime },l^{\prime })$ \ for \ $l^{\prime }\in
L_{k-1k-1}$.

Then the formula for the time-updated GLMB\ distribution at time \ $t_{k}$,
with spatial distributions \ $\mathring{s}_{l^{\prime },k|k-1}(x^{\prime })=%
\mathring{s}_{l^{\prime },k|k-1}(x^{\prime },l^{\prime })$, must be
specified: \ 
\begin{equation}
\mathring{f}_{k|k-1}(\mathring{X})=\delta _{|\mathring{X}|,|\mathring{X}_{%
\mathbb{L}}|}\,\sum_{o\in \mathbb{O}_{k|k-1}}\omega _{k|k-1}^{o}(\mathring{X}%
_{\mathbb{L}})\,(\mathring{s}_{k|k-1})^{\mathring{X}}.  \label{eq-GLMB-P}
\end{equation}%
\ 

The time-updated multitarget state \ $\mathring{X}$ \ at time \ $t_{k}$ \
can be partitioned as \ $\mathring{X}=\mathring{X}^{-}\uplus \mathring{X}%
^{+} $ where ??? 
\begin{eqnarray}
\mathring{X}^{-} &=&\mathring{X}\cap (\mathbb{X}_{0}\times L_{k-1|k-1})\text{
\ (surviving targets)} \\
\mathring{X}^{+} &=&\mathring{X}\cap (\mathbb{X}_{0}\times L_{k|k-1}^{B})%
\text{ \ \ \ \ (newly-appearing targets).}
\end{eqnarray}

The new-target distribution at time \ $t_{k}$ \ is assumed to be an LMB
distribution (as defined in Eq. (\ref{eq-LMB-1})),%
\begin{equation}
\mathring{b}_{k|k-1}(\mathring{X}^{+})=\delta _{|\mathring{X}^{+}|,|%
\mathring{X}_{\mathbb{L}}^{+}|}\,\omega _{k|k-1}^{B}(\mathring{X}_{\mathbb{L}%
}^{+})\,(\mathring{s}_{k|k-1}^{B})^{\mathring{X}^{+}},  \label{eq-Birth}
\end{equation}%
with, for some finite subset \ $L_{k|k-1}^{B}\subseteq \mathbb{L}_{k}$ \
(consisting of the labels of the hypothesized newly-appearing (a.k.a.
\textquotedblleft birth\textquotedblright ) targets), \ \ 
\begin{equation}
\omega _{k|k-1}^{B}(L)=\omega _{L_{k|k-1}^{B}}(L)=\left( \prod_{l\in
L_{k|k-1}^{B}-L}(1-q_{l}^{B})\right) \left( \prod_{l\in L}q_{l}^{B}\mathbf{1}%
_{L_{k|k-1}^{B}}(l)\right)  \label{eq-LMB}
\end{equation}%
and where \ $\mathring{s}_{l,k|k-1}^{B}(x)=\mathring{s}_{k|k-1}^{B}(x,l)$ \
with \ $l\in L_{k|k-1}^{B}$ \ are the corresponding spatial distributions.

Given this, the time-updated GLMB\ distribution is given by Eqs.
(15.277,15.281) of \cite{Mah-Newbook},%
\begin{eqnarray}
\omega _{k|k-1}^{o}(L) &=&\omega _{k|k-1}^{B}(L^{+})\,\tilde{\omega}%
_{k|k-1}^{o}(L^{-})  \label{eq-GLMB-P-w} \\
\mathring{s}_{k|k-1}(x,l) &=&\mathbf{1}_{L_{k|k-1}}(l)\,\mathring{s}%
_{k|k-1}^{o,S}(x,l)+\mathbf{1}_{L_{k|k-1}^{B}}(l)\,\mathring{s}%
_{k|k-1}^{B}(x,l),  \label{eq-GLMB-P-s}
\end{eqnarray}%
where a) \ $\mathbb{O}_{k|k-1}=\mathbb{O}_{k-1|k-1}$; b) \ $L^{+}\overset{_{%
\text{abbr.}}}{=}L\cap \mathbb{L}_{0:k-1}=L\cap \mathbb{L}_{k}=L\cap
L_{k|k-1}^{B}$ \ and \ $L^{-}\overset{_{\text{abbr.}}}{=}L\cap L_{k-1|k-1}$;
c) $\tilde{\omega}_{k|k-1}^{o}(J)$ \ is defined by \cite[Eq. (278)]%
{Mah-Newbook},\footnote{\textit{Errata}: \ The factor \ $\beta _{\mathring{X}%
_{\mathfrak{L}}^{\prime }}(\mathring{X}_{\mathfrak{L}}^{-})$ \ in \cite[Eq.
(15.286)]{Mah-Newbook} is redundent and therefore unnecessary. \ This is
because a)\ $\beta _{\mathring{X}_{\mathfrak{L}}^{\prime }}(\mathring{X}_{%
\mathfrak{L}}^{-})\in \{0,1\}$ $\ $and b) $\ \tilde{M}_{\mathring{X}^{-}}^{%
\mathring{X}^{\prime }}=0$ \ if \ $\beta _{\mathring{X}_{\mathfrak{L}%
}^{\prime }}(\mathring{X}_{\mathfrak{L}}^{-})=0$. \ Thus the \ $\beta
_{L}(J) $ \ in \cite[Eq. (15.278)]{Mah-Newbook} (the original version of Eq.
(\ref{Eq. (278)}) is not only unnecessary but erroneous, since in that case
\ $\sum_{o\in \mathbb{O}_{k|k-1}}\sum_{J\subseteq \mathbb{L}}\tilde{\omega}%
_{k|k-1}^{o}(J)\neq 1$. \ The factor \ $\mathbf{1}_{L}(l^{\prime })$ \ has
been included for clarity.} \ \ 
\begin{eqnarray}
&&\tilde{\omega}_{k|k-1}^{o}(J)  \label{Eq. (278)} \\
&=&\sum_{L\subseteq \mathbb{L}}\omega _{k-1|k-1}^{o}(L)\left(
\prod_{l^{\prime }\in L-J}\mathring{s}_{l^{\prime },k-1|k-1}^{o}[\mathring{p}%
_{S}^{c}]\right) \left( \prod_{l^{\prime }\in J}\mathbf{1}_{L}(l^{\prime })\,%
\mathring{s}_{l^{\prime },k-1|k-1}^{o}[\mathring{p}_{S}]\right) ;  \notag
\end{eqnarray}%
and d) the surviving-target spatial distributions are, for\ $\ l\in
L_{k-1|k-1}$, 
\begin{equation}
\mathring{s}_{l,k|k-1}^{o,S}(x)=\mathring{s}_{k|k-1}^{o,S}(x,l)=\frac{%
\mathring{s}_{l,k-1|k-1}^{o}[\mathring{p}_{S}\mathring{M}_{x}]}{\mathring{s}%
_{l,k-1|k-1}^{o}[\mathring{p}_{S}]}  \label{eq-GLMB-P-Big}
\end{equation}%
where%
\begin{equation}
\mathring{s}_{l,k-1|k-1}^{o}[\mathring{h}]=\int \mathring{h}(x,l)\,\mathring{%
s}_{k-1|k-1}^{o}(x,l)dx.
\end{equation}%
A clarified proof of Eq. (\ref{eq-GLMB-P-Big}) can be found in Section \ref%
{A-Der-AA-1}.

Written in expanded form, Eq. (\ref{eq-GLMB-P}) is \cite[Eq. (15.232)]%
{Mah-Newbook} \ 
\begin{eqnarray}
&&\mathring{f}_{k|k-1}(\mathring{X})  \label{eq-GLMB-P2} \\
&=&\delta _{|\mathring{X}|,|\mathring{X}_{\mathbb{L}}|}\,\omega _{k|k-1}^{B}(%
\mathring{X}_{\mathbb{L}}^{+})\sum_{o\in \mathbb{O}_{k|k-1}}\tilde{\omega}%
_{k|k-1}^{o}(\mathring{X}_{\mathbb{L}}^{-})\,(\mathring{s}_{k|k-1}^{B})^{%
\mathring{X}^{+}}\,(\mathring{s}_{k|k-1}^{o,S})^{\mathring{X}^{-}}.  \notag
\end{eqnarray}%
The time-updated spatial distributions are, therefore%
\begin{eqnarray}
\mathring{s}_{l,k|k-1}^{o,S}(x) &=&\mathring{s}_{k|k-1}^{o,S}(x,l)\text{ \ \
\ (}l\in L_{k-1|k-1}\text{, surviving targets)} \\
\mathring{s}_{l^{\prime },k|k-1}^{B}(x) &=&\mathring{s}_{k|k-1}^{B}(x,l^{%
\prime })\text{ \ \ \ (}l^{\prime }\in L_{k|k-1}^{B}\text{, birth targets).}
\end{eqnarray}

Note that, in Eq. (\ref{Eq. (278)}) and as is required for a GLMB\ p.d.f.,%
\begin{equation}
\sum_{o\in \mathbb{O}_{k|k-1}}\sum_{J\subseteq \mathbb{L}}\tilde{\omega}%
_{k|k-1}^{o}(J)=1.  \label{eq-Normal}
\end{equation}%
This is verified in Section \ref{A-Der-AA-2}.

\subsubsection{GLMB\ Filter Measurement-Update \label{A-Math-AA-GLMB-AAA-B}}

This subsection summarizes pp. 482-488 of \cite{Mah-Newbook}. \ 

For any measurement-set \ $Z$ \ with elements \ $z\in Z$, define \cite[Eq.
(15.187)]{Mah-Newbook}

\begin{equation}
\mathring{L}_{l,Z}^{\mathring{\theta}}(x)=\mathring{L}_{Z}^{\mathring{\theta}%
}(x,l)=\delta _{0,\mathring{\theta}(l)}\,\mathring{p}_{D}^{c}(x,l)+\delta
_{0,\mathring{\theta}(l)}^{c}\frac{\mathring{p}_{D}(x,l)\,\mathring{f}%
_{k}(z_{\mathring{\theta}(l)}|x,l)}{\kappa _{k}(z_{\mathring{\theta}(l)})}
\label{eq-GLMB-B-LIke}
\end{equation}%
where a) $\mathring{p}_{D}(x,l)$ \ is the probability of detection; b) \ $%
\mathring{p}_{D}^{c}(x,l)=1-\mathring{p}_{D}(x,l)$; c) \ $\delta _{0,%
\mathring{\theta}(l)}^{c}=1-\delta _{0,\mathring{\theta}(l)}$; \ d) \ $%
\mathring{f}_{k}(z|x,l)=\mathring{L}_{z}(x,l)$ \ is the single-target
likelihood function; e)$\ \kappa _{k}(z)$ \ is the intensity function of the
Poisson clutter process; and f) \ $\mathring{\theta}$ is a (labeled)
\textquotedblleft measurement-to-target association (MTA)\textquotedblright
---i.e., a function\ \ $\mathring{\theta}:L_{k|k-1}\rightarrow
\{0,1,...,|Z_{k}|\}$ \ such that $\mathring{\theta}(l)>0$ \ for only a
finite number of \ $l\in \mathbb{L}$ and such that \ $\mathring{\theta}(l)=%
\mathring{\theta}(l^{\prime })>0$ \ implies \ $l=l^{\prime }$. \ The set of
all MTAs at time \ $t_{k}$ \ is denoted \ $\mathcal{T}_{Z_{k}}$. \ \ 

Let us be given the predicted GLMB distribution at time \ $t_{k}$, \ 
\begin{equation}
\mathring{f}_{k|k-1}(\mathring{X})=\delta _{|\mathring{X}|,|\mathring{X}_{%
\mathbb{L}}|}\,\sum_{o\in \mathbb{O}_{k|k-1}}\omega _{k|k-1}^{o}(\mathring{X}%
_{\mathbb{L}})\,(\mathring{s}_{k|k-1})^{\mathring{X}},
\end{equation}%
where a)\ $\mathring{X}_{\mathbb{L}}\subseteq L_{k|k-1}$; b) the predicted
target spatial distributions are \ $\mathring{s}_{l,k|k-1}(x)=\mathring{s}%
_{k|k-1}(x,l)$; and c) the label-set\ is \ $L_{k|k}=L_{k|k-1}=L_{k|k-1}^{B}%
\uplus L_{k-1|k-1}\subseteq \mathbb{L}_{0:k}$. \ 

Assume that multitarget measurement-set \ $Z_{k}$ \ has been collected at
time \ $t_{k}$. \ Then\ the measurement-updated GLMB\ distribution is \cite[%
Eq. (15.229)]{Mah-Newbook} 
\begin{equation}
\mathring{f}_{k|k}(\mathring{X}|Z)=\delta _{|\mathring{X}|,|\mathring{X}_{%
\mathbb{L}}|}\sum_{(o,\mathring{\theta})\in \mathbb{O}_{k|k}}\omega
_{k|k}^{(o,\mathring{\theta})}(\mathring{X}_{\mathbb{L}})\,(\mathring{s}%
_{k|k}^{o,\mathring{\theta}})^{\mathring{X}}  \label{eq-GLMB-Meas-Init}
\end{equation}%
where: a) \ $\mathring{X}_{\mathbb{L}}\subseteq L_{k|k}$; b) \textit{\ }$%
\mathbb{O}_{k|k}=\mathbb{O}_{k|k-1}\times \mathcal{T}_{Z_{k}}=\mathcal{T}%
_{Z_{1}}\times ...\times \mathcal{T}_{Z_{k}}$; and thus\ c) the
measurement-updated weight function is, for \ $L\subseteq L_{k|k}$ \ and \ $%
(o,\mathring{\theta})\in \mathbb{O}_{k|k}$ \cite[Eq. (15.239)]{Mah-Newbook},
\ \ \ \ 
\begin{equation}
\omega _{k|k}^{(o,\mathring{\theta})}(L)=\frac{\omega _{k|k-1}^{o}(L)\,%
\mathring{\lambda}_{k}^{\mathring{\theta}}(L)\prod_{l\in L}\mathring{s}%
_{k|k-1}^{o,l}[\mathring{L}_{Z_{k}}^{\mathring{\theta}}]}{\sum_{L\subseteq 
\mathbb{L}}\sum_{\mathring{\theta}\in \mathcal{T}_{Z_{k}}}\sum_{o\in \mathbb{%
O}_{k|k-1}}\omega _{k|k-1}^{o}(L)\,\mathring{\lambda}_{k}^{\mathring{\theta}%
}(L)\prod_{l\in L}\mathring{s}_{k|k-1}^{o,l}[\mathring{L}_{Z_{k}}^{\mathring{%
\theta}}]};  \label{eq-GLMB-Meas-w}
\end{equation}%
with \cite[Eq. (15.243)]{Mah-Newbook} \ 
\begin{eqnarray}
\mathring{\lambda}_{k}^{\mathring{\theta}}(L) &=&\prod_{l\in
L_{k|k}-L}\delta _{\mathring{\theta}(l),0} \\
\mathring{s}_{k|k-1}^{o,l}[\mathring{L}_{Z_{k}}^{\mathring{\theta}}] &=&\int 
\mathring{L}_{Z_{k}}^{\mathring{\theta}}(x,l)\,\mathring{s}%
_{k|k-1}^{o}(x,l)dx;
\end{eqnarray}%
and d) the measurement-updated spatial distributions are, for \ $l\in
L_{k|k} $ \cite[Eq. (15.252)]{Mah-Newbook}, \ 

\begin{equation}
\mathring{s}_{l,k|k}^{(o,\mathring{\theta})}(x)=\frac{\mathring{s}%
_{l,k|k-1}^{o}(x)\,\mathring{L}_{l,Z_{k}}^{\mathring{\theta}}(x)}{\int 
\mathring{s}_{l,k|k-1}^{o}(x)\,\mathring{L}_{l,Z_{k}}^{\mathring{\theta}%
}(x)dx}.  \label{eq-GLMB-spatial}
\end{equation}

\subsection{Factorial Covariance \label{A-Math-AA-FacCov}}

The factorial covariance density of an LRFS \ $\mathring{\Xi}$ \ is \cite[%
Eq. (5.5.14d)]{Daley-Jones}, \cite[Eq. (21)]{Mah-YBS01}%
\begin{eqnarray}
\mathring{c}_{\mathring{\Xi}}^{[2]}(\mathring{x}_{1},\mathring{x}_{2}) &=&%
\frac{\delta ^{2}\log \mathring{G}_{\mathring{\Xi}}}{\delta \mathring{x}%
_{1}\delta \mathring{x}_{2}}[1]  \label{eq-FacCov} \\
&=&\frac{\delta ^{2}\mathring{G}_{\mathring{\Xi}}}{\delta \mathring{x}%
_{1}\delta \mathring{x}_{2}}[1]-\frac{\delta \mathring{G}_{\mathring{\Xi}}}{%
\delta \mathring{x}_{1}}[1]\frac{\delta \mathring{G}_{\mathring{\Xi}}}{%
\delta \mathring{x}_{2}}[1] \\
&=&\mathring{D}_{\mathring{\Xi}}(\mathring{x}_{1},\mathring{x}_{2})-%
\mathring{D}_{\mathring{\Xi}}(\mathring{x}_{1})\,\mathring{D}_{\mathring{\Xi}%
}(\mathring{x}_{2})
\end{eqnarray}%
where \ 
\begin{eqnarray}
\mathring{D}_{\mathring{\Xi}}(\mathring{x})\ &=&\frac{\delta \mathring{G}_{%
\mathring{\Xi}}}{\delta \mathring{x}}[1], \\
\mathring{D}_{\mathring{\Xi}}(\mathring{x}_{1},\mathring{x}_{2}) &=&\frac{%
\delta ^{2}\mathring{G}_{\mathring{\Xi}}}{\delta \mathring{x}_{1}\delta 
\mathring{x}_{2}}[1]
\end{eqnarray}%
are, respectively, the first-order and second-order factorial-moment density
functions of \ $\mathring{\Xi}$.\footnote{%
In the unlabeled case (i.e., \ $|\mathbb{L}|=1$), $\mathring{D}_{\mathring{%
\Xi}}(\mathring{x})$ \ is also known as the \textquotedblleft intensity
function\textquotedblright\ or \textquotedblleft probability hypothesis
density\textquotedblright\ (PHD of \ $\mathring{\Xi}$.}

As a simple example, suppose that $\ \mathring{\Xi}$ \ is an LMB RFS. \ Then
\ $\mathring{c}_{\mathring{\Xi}}^{[2]}(\mathring{x}_{1},\mathring{x}_{2})=0$
\ identically. \ This reflects the fact that \ $\mathring{\Xi}=\mathring{\Xi}%
_{1}\cup ....\cup \mathring{\Xi}_{n}$ \ where \ $\mathring{\Xi}_{1},...,%
\mathring{\Xi}_{n}$ \ are mutually independent Bernoulli LRFSs.

\section{Reformulated GLMB RFSs \label{A-Aug}}

The section is organized as follows: \ reformulated GLMB\ RFSs, first in
regard to their p.g.fl.'s (Section \ref{A-Aug-AA-PGFL}) and then their
p.d.f.'s (Section \ref{A-Aug-AA-PDF}); and the relevance of this
reformulation to multitarget statistical correlation (Section \ref%
{A-Aug-AA-Corr}).

\subsection{Reformulated GLMB:\ p.g.fl. \label{A-Aug-AA-PGFL}}

The definition of the p.g.fl. of a GLMB\ RFS, Eq. (\ref{eq-GLMB-PGFL}), can
be reformulated as follows: \ 
\begin{equation}
\mathring{G}_{\mathbb{L}}[\mathring{h}]=\sum_{L\subseteq \mathbb{L}}\omega
(L)\,\mathring{s}_{L}[\mathring{h}]  \label{eq-GLMB-Aug-1}
\end{equation}%
where the p.g.fl. \ $\mathring{s}_{L}[\mathring{h}]$ \ and weights \ $\alpha
_{o}^{L}$ \ and \ $\omega (L)$ \ are defined by \ \ \ 
\begin{eqnarray}
\mathring{s}_{L}[\mathring{h}] &=&\sum_{o\in \mathbb{O}}\alpha
_{o}^{L}\prod_{l\in L}\mathring{s}_{l}^{o}[\mathring{h}]
\label{eq-GLMB-Aug-2} \\
\alpha _{o}^{L} &=&\frac{\omega ^{o}(L)}{\omega (L)}  \label{eq-GLMB-Aug-3}
\\
\omega (L) &=&\sum_{o\in \mathbb{O}}\omega ^{o}(L)  \label{eq-GLMB-Aug-4}
\end{eqnarray}%
and which satisfy the following identities:%
\begin{eqnarray}
\sum_{o\in \mathbb{O}}\alpha _{o}^{L} &=&1\text{ \ \ \ \ \ \ (}L\subseteq 
\mathbb{L}\text{)}  \label{eq-GLMB-Aug-5} \\
\sum_{L\subseteq \mathbb{L}}\omega (L) &=&1.  \label{eq-GLMB-Aug-6}
\end{eqnarray}

Eq. (\ref{eq-GLMB-Aug-1}) is true because, by Eq. (\ref{eq-GLMB-PGFL}), \ \
\ 
\begin{eqnarray}
\mathring{G}_{\mathbb{L}}[\mathring{h}] &=&\sum_{o\in \mathbb{O}%
}\sum_{L\subseteq \mathbb{L}}\omega ^{o}(L)\prod_{l\in L}\mathring{s}%
_{l}^{o}[\mathring{h}] \\
&=&\sum_{L\subseteq \mathbb{L}}\omega (L)\sum_{o\in \mathbb{O}}\alpha
_{o}^{L}\prod_{l\in L}\mathring{s}_{l}^{o}[\mathring{h}] \\
&=&\sum_{L\subseteq \mathbb{L}}\omega (L)\,\mathring{s}_{L}[\mathring{h}].
\end{eqnarray}

\subsection{Reformulated GLMB:\ p.d.f. \label{A-Aug-AA-PDF}}

The p.d.f. corresponding to Eq. (\ref{eq-GLMB-Aug-1}) is \ \ \ 
\begin{equation}
\mathring{f}_{\mathbb{L}}(\mathring{X})=\delta _{|\mathring{X}|,|\mathring{X}%
_{\mathbb{L}}|}\,\omega (\mathring{X}_{\mathbb{L}})\,\mathring{s}_{\mathring{%
X}_{\mathbb{L}}}(\vec{x}_{L}),  \label{eq-GLMB-Aug-PDF-1}
\end{equation}%
where the vector \ $\vec{x}_{L}$ \ will be defined shortly in Eq. (\ref%
{eq-00}) and where%
\begin{equation}
\mathring{s}_{L}(\vec{x}_{L})=\sum_{o\in \mathbb{O}}\alpha
_{o}^{L}\prod_{l\in L}\mathring{s}_{l}^{o}(\mathring{X}_{l})\text{ \ \ \ \ \
\ \ \ \ (}L\subseteq \mathbb{L}\text{)},  \label{eq-GLMB-Aug-PDF-2}
\end{equation}%
with \ $\mathring{X}_{l}$ \ as defined in Eq. (\ref{eq-Big}).

More specifically, let \ $\mathring{X}=\{(x_{1},l_{1}),...,(x_{n},,l_{n})\}$
\ with \ $|\mathring{X}|=|\mathring{X}_{\mathbb{L}}|=n$. \ Then%
\begin{equation}
\mathring{s}_{\{l_{1},...,l_{n}\}}(x_{1},...,x_{n})=\sum_{o\in \mathbb{O}%
}\alpha _{o}^{\{l_{1},...,l_{n}\}}\prod_{i=1}^{n}\mathring{s}%
_{l_{i}}^{o}(x_{i}).
\end{equation}%
Note that \ $\mathring{s}_{\{l_{1},...,l_{n}\}}(x_{1},...,x_{n})$ \ is not
symmetric in the arguments \ $\ x_{1},...,x_{n}$.

\begin{remark}
\label{Rem-GenLMB}Comparing Eqs. (\ref{eq-GLMB-Aug-PDF-1},\ref%
{eq-GLMB-Aug-PDF-2}) to Eq. (\ref{eq-LMB-1}), note that GLMB\ p.d.f.'s can
be interpreted as direct generalizations of LMB p.d.f.'s to correlated
targets, with \ $\omega _{J}(\mathring{X}_{\mathbb{L}})$ \ replaced by \ $%
\omega (\mathring{X}_{\mathbb{L}})$ \ and \ $\prod_{i=1}^{n}\mathring{s}%
_{l_{i}}(x_{i})$ by \ $\sum_{o\in \mathbb{O}}\alpha _{o}^{\mathring{X}_{%
\mathbb{L}}}\prod_{i=1}^{n}\mathring{s}_{l_{i}}^{o}(x_{i})$.
\end{remark}

It follows that, for each \ $n$, the marginal distribution for labels, 
\begin{equation}
\int \mathring{f}_{\mathbb{L}}((x_{1},l_{1}),...,(x_{n},,\NEG%
{l}_{n}))dx_{1}\cdots dx_{n}=\omega (\{l_{1},...,l_{n}\}),
\end{equation}%
is the probability\footnote{$\omega (L)$ is a \textquotedblleft basic mass
assignment,\textquotedblright\ in the terminology of Dempster-Shafer theory}
that the targets in the multitarget population are those with distinct
labels \ $l_{1},...,l_{n}$.

Eq. (\ref{eq-GLMB-Aug-PDF-1}) follows from \ 
\begin{eqnarray}
\mathring{f}(\mathring{X}) &=&\delta _{|\mathring{X}|,|\mathring{X}_{\mathbb{%
L}}|}\sum_{o\in \mathbb{O}}\omega ^{o}(\mathring{X}_{\mathbb{L}})\,(%
\mathring{s}^{o})^{\mathring{X}} \\
&=&\delta _{|\mathring{X}|,|\mathring{X}_{\mathbb{L}}|}\,\omega (\mathring{X}%
_{\mathbb{L}})\sum_{o\in \mathbb{O}}\alpha _{o}^{\mathring{X}_{\mathbb{L}%
}}\,(\mathring{s}^{o})^{\mathring{X}} \\
&=&\delta _{|\mathring{X}|,|\mathring{X}_{\mathbb{L}}|}\,\omega (\mathring{X}%
_{\mathbb{L}})\sum_{o\in \mathbb{O}}\alpha _{o}^{\mathring{X}_{\mathbb{L}%
}}\prod_{l\in \mathring{X}_{\mathbb{L}}}\mathring{s}_{l}^{o}(\mathring{X}%
_{l}) \\
&=&\delta _{|\mathring{X}|,|\mathring{X}_{\mathbb{L}}|}\,\omega (\mathring{X}%
_{\mathbb{L}})\,\mathring{s}_{\mathring{X}_{\mathbb{L}}}(\vec{x}_{L}).
\end{eqnarray}

\subsection{Reformulated GLMBs:\ Correlation \label{A-Aug-AA-Corr}}

Note that \ $\mathring{s}_{L}(\vec{x}_{L})$ \ is a p.d.f. on \ $\mathbb{X}%
_{0}^{|L|}$ \ (i.e., its integral is $1$) and that its $\ n$ \ marginal
p.d.f.'s are, for \ $l\in L$, \ \ \ 
\begin{equation}
\mathring{s}_{l|L}(x)\overset{_{\text{def.}}}{=}\sum_{o\in \mathbb{O}}\alpha
_{o}^{L}\,\mathring{s}_{l}^{o}(x).  \label{eq-GLMB-Aug-Marg}
\end{equation}

In general, it will not be the case that \ $\mathring{s}_{l|L}^{o}(x)$ \
factorizes into its marginals, i.e., that 
\begin{equation}
\mathring{s}_{\{l_{1},...,l_{n}\}}(x_{1},...,x_{n})=\mathring{s}%
_{l_{1}|L}(x_{1})\cdots \mathring{s}_{l_{n}|L}(x_{n})
\end{equation}%
for all \ $x_{1},...,x_{n}$. \ Thus \ $\mathring{s}_{\{l_{1},...,l_{n}%
\}}(x_{1},...,x_{n})$ \ is, implicitly, a model of the statistical
correlation of the targets whose labels are in \ $\{l_{1},...,l_{n}\}$%
---what in Section \ref{A-SLC} will be called a \textquotedblleft simple
labeled correlation (SLC) model.\textquotedblright\ 

\section{Simple Labeled Correlation Models \label{A-SLC}}

In GLMB\ filter theory, it is assumed that at any given time there is a) a
finite set \ $\mathbb{L}$ \ of labels, and b) an indexed family \ $\{%
\mathring{s}_{l}\}_{l\in \mathbb{L}}$ \ of target spatial distributions \ $%
\mathring{s}_{l}(x)=\mathring{s}(x,l)$. \ This models a system of targets
indexed by \ $l\in \mathbb{L}$\ \ that are, initially, mutually
statistically independent. \ This section addresses targets that are,
instead, initially assumed to be possibly statistically dependent.

It is organized as follows: \ SLC\ models (Section \ref{a-SLC-AA-Def});
interpretation of SLC\ models (Section \ref{a-SLC-AA-Interp}); and the
relationship between GLMB RFSs and SLC\ models (Section \ref%
{a-SLC-AA-Example}).

\subsection{SLC Models \ \label{a-SLC-AA-Def}}

For every \ $L\subseteq \mathbb{L}$, we are given:

\begin{enumerate}
\item finite index-sets \ $\mathbb{I}_{L}$ \ with corresponding indices \ $%
i_{L}\in \mathbb{I}_{L}$;

\item for each \ $l\in L$,\ a family \ $\mathring{s}_{l}^{i_{L}}(x)$ \ of
single-target spatial p.d.f.'s on \ $\mathbb{X}_{0}$, indexed by \ $i_{L}$;

\item weights $\ 0\leq \alpha _{i_{L}}^{L}\leq 1$ \ such that \ $%
\sum_{i_{L}\in \mathbb{I}_{L}}\alpha _{i_{L}}^{L}=1$; and \ 

\item for notational convenience, an ordering on the elements of \ $\mathbb{L%
}$, in which case $\ $%
\begin{equation}
\vec{x}_{L}=(x_{l})_{l\in L}\in \mathbb{X}_{0}^{|L|}  \label{eq-00}
\end{equation}%
can be regarded as a vector defined by concatenation.\footnote{%
An ordering \ $\mathring{X}=\{\mathring{x}_{1},...,\mathring{x}_{n}\}$ \ of
the elements \ $\mathring{x}_{1},...,\mathring{x}_{n}$ \ of an LFS \ $%
\mathring{X}$ \ \ is a choice of notational convenience, not mathematical
necessity. \ The same is true of the implicit ordering in Eq. (\ref{eq-00})
of the elements in \ $\mathbb{L}$ \ and thus also \ $\vec{x}_{L}$.}
\end{enumerate}

Such systems \ 
\begin{equation}
\{\mathbb{I}_{L},\mathring{s}_{l}^{i_{L}},\alpha _{i_{L}}^{L}\}_{L\subseteq 
\mathbb{L},l\in L,i_{L}\in \mathbb{I}_{L}}
\end{equation}%
will henceforth be called \textit{simple labeled correlation }(SLC)\textit{\
models}.\ \ 

SLC representation includes GLMB representation as a special case. \ For, if
\ $L=\{l\}$ for all \ $l\in \mathbb{L}$ \ then the former reduces to \ $\{%
\mathbb{I}_{\{l\}},\mathring{s}_{l}^{i_{\{l\}}},\alpha
_{i_{\{l\}}}^{\{l\}}\}_{l\in \mathbb{L},i_{\{l\}\in \mathbb{I}_{\{l\}}}}$ \
and thereby to \ $\{\mathring{s}_{l|\{l\}}\}_{l\in \mathbb{L}}$\ \ where \ 
\begin{equation}
\mathring{s}_{l|\{l\}}(x)\overset{_{\text{def.}}}{=}\sum_{i_{\{l\}}\in 
\mathbb{I}_{\{l\}}}\alpha _{i_{\{l\}}}^{\{l\}}\,\mathring{s}%
_{l}^{i_{\{l\}}}(x).
\end{equation}

\subsection{Interpretation of SLC Models \label{a-SLC-AA-Interp}}

For each \ $L\subseteq \mathbb{L}$ \ and each \ $l\in L$ \ and each \ $%
i_{L}\in \mathbb{I}_{L}$, \ $\mathring{s}_{l_{L}}^{i_{L}}(x)$ \ can be
regarded as a hypothesized spatial p.d.f. for the target with label \ $l$ \
when \ $l$ \ is considered as a member of the target-cluster \ $L$. \ 

This interpretation is justifiable as follows. \ When the targets \ $l\in L$
\ in the target cluster \ $L$ \ are very closely spaced (and thus very
difficult to distinguish from each other), the choice of the hypothesized
spatial p.d.f. \ $\mathring{s}_{l}(x)$\ of \ $l$ \ is rather ambiguous. \ It
is therefore reasonable to hedge against this uncertainty by modeling it\ as
the discrete random p.d.f. \ $\mathbf{\mathring{s}}_{l}^{i_{L}}(x)$ \
defined by \ \ 
\begin{equation}
\Pr (\mathbf{\mathring{s}}_{l}^{i_{L}}=\mathring{s}_{l}^{i_{L}})=\alpha
_{i_{L}}^{L}\ \ \ \ \text{(}i_{L}\in \mathbb{I}_{L}\text{).}
\label{eq-Random}
\end{equation}

Given this, the expected value of \ $\mathbf{\mathring{s}}_{l}^{i_{L}}$,%
\begin{equation}
\mathring{s}_{l|L}(x_{l})\overset{_{\text{def.}}}{=}\sum_{i_{L}\in \mathbb{I}%
_{L}}\alpha _{i_{L}}^{L}\,\mathring{s}_{l}^{i_{L}}(x_{l}),
\end{equation}%
can be regarded as an estimate of target \ $l$'s actual spatial p.d.f. when
\ $l$ \ is regarded as a member of the target-cluster \ $L$. \ 

Similarly, for each \ $L$ \ and each \ $i_{L}\in \mathbb{I}_{L}$, the p.d.f.
\ $\prod_{l\in L}\mathring{s}_{l}^{i_{L}}(x_{l})\ \ $(where \ $x_{l}\in 
\mathbb{X}_{0}$ \ with \ $l\in L$ \ denotes the \ $l$'th component of \ $%
\vec{x}_{L}$) can be regarded as a hypothesis that the targets in \ $L$ \
are statistically independent. \ Given this, the p.d.f. \ \ 
\begin{equation}
\mathring{s}_{L}(\vec{x}_{L})=\sum_{i_{L}\in \mathbb{I}_{L}}\alpha
_{i_{L}}^{L}\prod_{l\in L}\mathring{s}_{l}^{i_{L}}(x_{l})
\end{equation}%
can be regarded as an estimate of the joint distribution of the \ $l\in L$ \
since its \ $|L|$\ \ marginal p.d.f.'s are the \ $\mathring{s}_{l|L}(x_{l})$
\ for \ $l\in L$.\ 

In general it will not be true that%
\begin{equation}
\mathring{s}_{L}(\vec{x}_{L})=\prod_{l\in L}\mathring{s}_{l|L}(x_{l})
\end{equation}%
for all \ $\vec{x}_{L}$i.e., that the targets in \ $L$ \ are statistically
independent. \ It follows that SLC models---and GLMB\ distributions in
particular---are algebraically simple models of multitarget correlation. \ 

The degree of correlation can be measured using the factorial covariance
density (f.c.d.) \ $\mathring{c}_{\mathring{\Xi}}^{[2]}(\mathring{x}_{1},%
\mathring{x}_{2})$, \ as defined in Eq. (\ref{eq-FacCov}). \ For example,
consider the special case of Eq. (\ref{eq-GLMB-Aug-1}) with \ $|\mathbb{O}%
|=1 $ \ and \ $\omega (L)=\delta _{L,\{l_{1},l_{2}\}}$ \ and \ $l_{1}\neq
l_{2}$: \ 
\begin{equation}
\mathring{G}_{\mathbb{L}}[\mathring{h}]=\mathring{s}_{\{l_{1},l_{2}\}}[%
\mathring{h}]=\sum_{o\in \mathbb{O}}\alpha _{o}^{\{l_{1},l_{2}\}}\,\mathring{%
s}_{l_{1}}^{o}[\mathring{h}]\,\mathring{s}_{l_{2}}^{o}[\mathring{h}].
\end{equation}%
If \ $\mathring{x}_{1}^{\prime }=(x_{1}^{\prime },l_{1}^{\prime })$ \ and \ $%
\mathring{x}_{2}^{\prime }=(x_{2}^{\prime },l_{2}^{\prime })$ \ then the
corresponding f.c.d.\ is: \ 
\begin{eqnarray}
&&\mathring{c}_{\{l_{1},l_{2}\}}^{[2]}(\mathring{x}_{1}^{\prime },\mathring{x%
}_{2}^{\prime })  \label{eq-FacCov-2} \\
&=&\delta _{\{l_{1},l_{2}\},\{l_{1}^{\prime },l_{2}^{\prime }\}}\left( 
\mathring{s}_{\{l_{1}^{\prime },l_{2}^{\prime }\}}(x_{1}^{\prime
},x_{2}^{\prime })-\mathring{s}_{l_{1}^{\prime }|\{l_{1}^{\prime
},l_{2}^{\prime }\}}(x_{1}^{\prime })\,\mathring{s}_{l_{2}^{\prime
}|\{l_{1}^{\prime },l_{2}^{\prime }\}}(x_{2}^{\prime })\right) .  \notag
\end{eqnarray}%
This is verifed in Section \ref{A-Der-AA-2a}.

\subsection{SLC Interpretation of GLMB \label{a-SLC-AA-Example}}

This section provides a detailed illustration of the relationship between
SLC\ models and the GLMB\ filter. \ 

Consider the GLMB\ filter at times \ $t_{0}$ \ and \ $t_{1}$ \ with \ $%
\mathring{f}_{0|0}(\emptyset )=1$ \ (i.e., the initial scenario is
target-free). \ Then a)$\ \ L_{1|0}=L_{1|0}^{B}\subseteq \mathbb{L}_{0:1}$ \
is the label-set for the initial set of birth targets; b) a measurement-set
\ $Z_{1}$ \ is collected at time $t_{1}$; and c) \ $\mathbb{O}_{1|1}=%
\mathcal{T}_{Z_{1}}$\ \ where $\mathcal{T}_{Z_{1}}$ \ is the set of MTAs \ $%
\mathring{\theta}:L_{1}^{B}\rightarrow \{0,1,...,|Z_{1}|\}$. \ 

Given this, the measurement-updated SLC-GLMB\ p.d.f. is%
\begin{equation}
\mathring{f}_{1|1}(\mathring{X})=\delta _{|\mathring{X}|,|\mathring{X}_{%
\mathbb{L}}|}\,\omega _{1|1}(\mathring{X}_{\mathbb{L}})\sum_{\mathring{\theta%
}\in \mathcal{T}_{Z_{1}}}\alpha _{\mathring{\theta},1|1}^{\mathring{X}_{%
\mathbb{L}}}\prod_{l\in \mathring{X}_{\mathbb{L}}}\mathring{s}_{l}^{%
\mathring{\theta}}(\mathring{X}_{l})\,
\end{equation}%
where \ $\mathring{X}_{l}$ \ was defined in Eq. (\ref{eq-Big}), where \ 
\begin{eqnarray}
\omega _{1|1}(L) &=&\sum_{\mathring{\theta}\in \mathcal{T}_{Z_{1}}}\omega
_{1|1}^{\mathring{\theta}}(L) \\
\alpha _{\mathring{\theta},1|1}^{L} &=&\frac{\omega _{1|1}^{\mathring{\theta}%
}(L)}{\omega _{1|1}(L)}\text{ \ \ \ \ \ \ \ \ \ \ \ \ (}L\subseteq L_{1}^{B}%
\text{)} \\
\mathring{s}_{l,1:1}^{\mathring{\theta}}(x) &\propto &\mathring{s}%
_{l,1|0}^{o}(x)\,\mathring{L}_{l,Z_{1}}^{\mathring{\theta}}(x)\text{ \ \ (}%
l\in L_{1}^{B}\text{),}
\end{eqnarray}%
where \ $\omega _{1|1}^{\mathring{\theta}}(\mathring{X}_{\mathbb{L}})$ \ is
as defined as in Eq. (\ref{eq-GLMB-Meas-w}),\ and where \ $\mathring{s}%
_{l,1:1}^{\mathring{\theta}}(x)$ \ is as in Eq. (\ref{eq-GLMB-spatial}).

Thus the joint\ p.d.f.'s are, for \ $L\subseteq L_{1|0}$ \ (and for \ $%
\mathring{X}_{l}$ \ as defined in Eq. (\ref{eq-Big})) 
\begin{equation}
\mathring{s}_{L}^{1|1}(x_{1},...,x_{|L|})=\sum_{\mathring{\theta}\in 
\mathcal{T}_{Z_{1}}}\alpha _{\mathring{\theta}}^{L}\prod_{l\in L}\mathring{s}%
_{l,1|1}^{\mathring{\theta}}(\mathring{X}_{l})\text{.}
\end{equation}%
The $|L|$ \ marginal p.d.f.'s of this are 
\begin{equation}
\mathring{s}_{l|L}^{1|1}(x)=\sum_{\mathring{\theta}\in \mathcal{T}%
_{Z_{1}}}\alpha _{\mathring{\theta}}^{L}\,\mathring{s}_{l,1|1}^{\mathring{%
\theta}}(x)\text{ \ \ \ \ (}l\in L\text{).}  \label{eq-estimate}
\end{equation}%
Eq. (\ref{eq-estimate}) is an estimate of the spatial distribution of \ $l$
\ when regarded as part of the target-cluster \ $L$. \ As such, it is a
weighted average of the hypothesized spatial distribution \ $\mathring{s}%
_{l,1|1}^{\mathring{\theta}}(x)$ \ for all MTAs \ $\mathring{\theta}\in 
\mathcal{T}_{Z_{1}}$.

\section{The SLC-GLMB Filter \label{A-SLCGLMB}}

The GLMB\ filter has the algebraically exact-closed form%
\begin{equation}
...\rightarrow \left\{ 
\begin{array}{c}
\mathbb{O}_{k-1|k-1} \\ 
\omega _{k-1|k-1}^{o}(L) \\ 
\mathring{s}_{k-1|k-1}^{o}(\mathring{x})%
\end{array}%
\right. \rightarrow \left\{ 
\begin{array}{c}
\mathbb{O}_{k|k-1} \\ 
\omega _{k|k-1}^{o}(L) \\ 
\mathring{s}_{k|k-1}^{o}(\mathring{x})%
\end{array}%
\right. \rightarrow \left\{ 
\begin{array}{c}
\mathbb{O}_{k|k} \\ 
\omega _{k|k}^{o}(L) \\ 
\mathring{s}_{k|k}^{o}(\mathring{x})%
\end{array}%
\right. \rightarrow ...
\end{equation}%
Specifically, the triple \ $\mathbb{O}$, $\omega ^{o}(L)$, $\mathring{s}^{o}(%
\mathring{x})$ \ in \ each filtering step is recursively defined in terms of
the triple \ $\mathbb{O}$, $\omega ^{o}(L)$, $\mathring{s}^{o}(\mathring{x})$
\ \ in the filtering step that precedes it. \ 

The SLC-GLMB filter is also algebraically exact-closed:%
\begin{equation}
...\rightarrow \left\{ 
\begin{array}{c}
\mathbb{O}_{k-1|k-1} \\ 
\omega _{k-1|k-1}(L) \\ 
\alpha _{o,k-1|k-1}^{L} \\ 
\mathring{s}_{k-1|k-1}^{o}(\mathring{x})%
\end{array}%
\right. \rightarrow \left\{ 
\begin{array}{c}
\mathbb{O}_{k|k-1} \\ 
\omega _{k|k-1}(L) \\ 
\alpha _{o,k|k-1}^{L} \\ 
\mathring{s}_{k|k-1}^{o}(\mathring{x})%
\end{array}%
\right. \rightarrow \left\{ 
\begin{array}{c}
\mathbb{O}_{k|k} \\ 
\omega _{k|k}(L) \\ 
\alpha _{o,k|k}^{L} \\ 
\mathring{s}_{k|k}^{o}(\mathring{x})%
\end{array}%
\right. \rightarrow ...
\end{equation}%
Specifically, the quadruple \ $\mathbb{O}$, $\omega (L)$, $\alpha _{o}^{L}$, 
$\mathring{s}^{o}(\mathring{x})$ \ in \ each filtering step is recursively
defined purely in terms of the quadruple \ $\mathbb{O}$, $\omega (L)$, $%
\alpha _{o}^{L}$, $\mathring{s}^{o}(\mathring{x})$ \ in the filtering step
that precedes it. \ That is, \ $\mathbb{O}$, $\omega (L)$, $\alpha _{o}^{L}$%
, $\mathring{s}^{o}(\mathring{x})$ \ does not \textit{explicitly} depend on
the triple \ $\mathbb{O}$, $\omega ^{o}(L)$, $\mathring{s}^{o}(\mathring{x})$
\ that was originally used to define it.

The purpose of this section is to derive explicit formulas for the SLC-GLMB
filter. \ It is organized as follows: \ the SLC-GLMB\ filter time-update
(Section \ref{A-SLCGLMB-AA-P}); the SLC-GLMB\ filter measurement-update
(Section \ref{A-SLCGLMB-AA-B}); and SLC-GLMB\ filter state estimation
(Section \ref{A-SLCGLMB-AA-E}).

\subsection{SLC-GLMB Filter: \ Time-Update \label{A-SLCGLMB-AA-P}}

SLC-GLMB p.d.f.'s are exact-closed-form with respect to time-updates. \ That
is, for \ $\mathbb{O}_{k|k-1}=\mathbb{O}_{k-1|k-1}$ \ and \ $L\subseteq
L_{k|k-1}=L_{k|k-1}^{B}\uplus L_{k-1|k-1}$, \ \ 
\begin{eqnarray}
&&\omega _{k|k-1}(L)  \label{eq-SLC-P-w} \\
&=&\omega _{k|k-1}^{B}(L^{+})\sum_{L\subseteq L_{k-1|k-1}}\omega
_{k-1|k-1}(L)\sum_{o\in \mathbb{O}_{k|k-1}}\omega
_{k|k-1}^{S,o}(L^{-}|L)\,\alpha _{o,k-1|k-1}^{L}  \notag
\end{eqnarray}%
\begin{eqnarray}
\alpha _{o,k|k-1}^{L} &=&\alpha _{o,k|k-1}^{L^{-}}=\sum_{L\subseteq
L_{k-1|k-1}}\sigma _{k|k-1}^{S,o}(L^{-}|L)\,\alpha _{o,k-1|k-1}^{L}
\label{eq-SLC-P-aa} \\
&\propto &\sum_{L\subseteq L_{k-1|k-1}}\omega
_{k|k-1}^{S,o}(L^{-}|L)\,\omega _{k-1|k-1}(L)\,\alpha _{o,k-1|k-1}^{L} \\
\mathring{s}_{k|k-1}^{o}(x,l) &=&\mathbf{1}_{L_{k|k-1}}(l)\,\mathring{s}%
_{k|k-1}^{o,S}(x,l)+\mathbf{1}_{L_{k|k-1}^{B}}(l)\,\mathring{s}%
_{k|k-1}^{B}(x,l)  \label{eq-SLC-P-ss}
\end{eqnarray}%
where a) \ $\omega _{k|k-1}^{B}(L)$ \ was defined in Eq. (\ref{eq-LMB}); b)\ 
$\ \mathring{s}_{k|k-1}^{S,o}(x,l)$ \ was defined in Eq. (\ref{eq-GLMB-P-Big}%
; and c) for \ $L^{-},L\subseteq L_{k-1|k-1}$, \ 
\begin{eqnarray}
&&\omega _{k|k-1}^{S,o}(L^{-}|L)  \label{eq-SLC-P-t} \\
&=&\left( \prod_{l^{\prime }\in L-L^{-}}\mathring{s}_{l^{\prime
},k-1|k-1}^{o}[\mathring{p}_{S}^{c}]\right) \left( \prod_{l^{\prime }\in
L^{-}}\mathbf{1}_{L}(l^{\prime })\,\mathring{s}_{l^{\prime },k-1|k-1}^{o}[%
\mathring{p}_{S}]\right)   \notag \\
&&\sigma _{k|k-1}^{S,o}(L^{-}|L) \\
&=&\frac{\omega _{k|k-1}^{S,o}(L^{-}|L)\,\omega _{k-1|k-1}(L)}{%
\sum_{L\subseteq \mathbb{L}}\omega _{k-1|k-1}(L)\sum_{o\in \mathbb{O}%
_{k|k-1}}\,\omega _{k|k-1}^{S,o}(L^{-}|L)\,\alpha _{o,k-1|k-1}^{L}}.  \notag
\end{eqnarray}%
Note that \ $\omega _{k|k-1}^{S,o}(L^{-}|L)=0$ \ unless \ $L^{-}\subseteq L$
\ and that \ 
\begin{equation}
\sum_{L^{-}:L^{-}\subseteq L}\omega _{k|k-1}^{S,o}(L^{-}|L)=1.
\end{equation}

Eqs. (\ref{eq-SLC-P-w},\ref{eq-SLC-P-aa}) are derived in Section \ref%
{A-Der-AA-3}.

\subsection{SLC-GLMB Filter: \ Measurement-Update \label{A-SLCGLMB-AA-B}}

SLC-GLMB p.d.f.'s are exact-closed-form with respect to measurement-updates.
\ That is, for \ $\mathbb{O}_{k|k}=\mathbb{O}_{k|k-1}\times \mathcal{T}%
_{Z_{k}}$ \ and $L\subseteq L_{k|k}=L_{k|k-1}$, \ 
\begin{equation}
\omega _{k|k}(L)=\omega _{k|k-1}(L)\,\sum_{o\in \mathbb{O}_{k|k-1}}\alpha
_{o,k|k-1}^{L}\sum_{\mathring{\theta}\in \mathcal{T}_{Z_{k}}}\rho _{k|k}^{(o,%
\mathring{\theta})}(Z_{k}|L)  \label{eq-SLC-B-w}
\end{equation}%
\begin{eqnarray}
\alpha _{(o,\mathring{\theta}),k|k}^{L} &=&\frac{\alpha _{o,k|k-1}^{L}\,\rho
_{k|k}^{(o,\mathring{\theta})}(Z_{k}|L)}{\sum_{o\in \mathbb{O}%
_{k|k-1}}\alpha _{o,k|k-1}^{L}\sum_{\mathring{\theta}\in \mathcal{T}%
_{Z_{k}}}\rho _{k|k}^{(o,\mathring{\theta})}(Z_{k}|L)}  \label{eq-SLC-B-aa}
\\
&\propto &\mathring{\lambda}_{k}^{\mathring{\theta}}(L)\prod_{l\in L}%
\mathring{s}_{l,k|k-1}^{o}[\mathring{L}_{Z_{k}}^{\mathring{\theta}}] \\
\mathring{s}_{k|k}^{(o,\mathring{\theta}),S}(x,l) &=&\frac{\mathring{s}%
_{k|k-1}^{o}(x,l)\,\mathring{L}_{Z_{k}}^{\mathring{\theta}}(x,l)}{\int 
\mathring{s}_{k|k-1}^{o}(x,l)\,\mathring{L}_{Z_{k}}^{\mathring{\theta}%
}(x,l)dx}  \label{eq-SLC-B-ss}
\end{eqnarray}%
where \ $\mathring{L}_{Z_{k}}^{\mathring{\theta}}(x,l)$ \ was defined in Eq.
(\ref{eq-GLMB-B-LIke}) and where%
\begin{equation}
\rho _{k|k}^{(o,\mathring{\theta})}(Z_{k}|L)=\frac{\mathring{\lambda}_{k}^{%
\mathring{\theta}}(L)\prod_{l\in L}\mathring{s}_{l,k|k-1}^{o}[\mathring{L}%
_{Z_{k}}^{\mathring{\theta}}]}{\left( 
\begin{array}{c}
\sum_{L\subseteq L_{k|k-1}}\omega _{k|k-1}(L)\,\sum_{o\in \mathbb{O}%
_{k|k-1}}\alpha _{o,k|k-1}^{L}\, \\ 
\cdot \sum_{\mathring{\theta}\in \mathcal{T}_{Z_{k}}}\mathring{\lambda}_{k}^{%
\mathring{\theta}}(L)\,\omega _{k|k-1}^{o}(L)\prod_{l\in L}\mathring{s}%
_{l,k|k-1}^{o}[\mathring{L}_{Z_{k}}^{\mathring{\theta}}]%
\end{array}%
\right) }.
\end{equation}

Eqs. (\ref{eq-SLC-B-w},\ref{eq-SLC-B-aa}) are demonstrated in Section \ref%
{A-Der--AA-4}.

\subsection{SLC-GLMB Filter: \ State Estimation \label{A-SLCGLMB-AA-E}}

Eq. (\ref{eq-GLMB-Aug-PDF-1}) suggests the following multitarget
state-estimation approach for the SLC GLMB filter:

\begin{enumerate}
\item Determine the most probable target-set,%
\begin{equation}
\hat{L}_{k|k}=\arg \max_{L\in \mathbb{L}}\,\omega _{k|l}(L),
\end{equation}%
along with the most probable target-number \ $\hat{n}_{k|k}=|\hat{L}_{k|k}|$%
. \ 

\item The corresponding marginal p.d.f.'s of the joint distribution, as
defined in Eq. (\ref{eq-GLMB-Aug-Marg}), are the estimated target spatial
distributions:%
\begin{equation}
\hat{s}_{\hat{L}}^{k|k}(x,l)=\sum_{o\in \mathbb{O}}\alpha _{o,k|k}^{\hat{L}%
}\,\mathring{s}_{k|k}^{o}(x,l)\text{ \ \ \ \ \ (}l\in \hat{L}_{k|k}\text{)}.
\end{equation}%
\ \ \ 

\item The estimated target states are%
\begin{equation}
\mathring{x}_{l|\hat{L}}^{k|k}=\arg \sup_{x\in \mathbb{X}_{0}}\,\hat{s}_{l|%
\hat{L}}^{k|k}(x)\text{ \ \ \ \ \ \ \ (}l\in \hat{L}_{k|k}\text{)}
\label{eq-GLMB-Faster}
\end{equation}
\end{enumerate}

An approximation of Eq. (\ref{eq-GLMB-Faster})---one which resembles the
state-estimation procedure employed in multi-hypothesis tracking (MHT)---is
to determine the most probable \ $o$ \ for \ $\hat{L}_{k|k}$, \ 
\begin{equation}
\hat{o}_{\hat{L}}^{k|k}=\arg \max_{o\in \mathbb{O}_{k|k}}\,\alpha _{o}^{\hat{%
L}_{k|k}}\text{ =}\arg \max_{o\in \mathbb{O}_{k|k}}\,\omega ^{o}(\hat{L}%
_{k|k})\text{,}  \label{eq-Aug-AA-E-Approx1}
\end{equation}%
and then determine \ 
\begin{equation}
\mathring{x}_{l|\hat{L}_{k|k}}^{k|k}=\arg \sup_{x\in \mathbb{X}_{0}}\,%
\mathring{s}_{k|k}^{\hat{o}_{\hat{L}_{k|k}}}(x,l)\text{ \ \ \ \ \ \ \ (}l\in 
\hat{L}_{k|k}\text{).}  \label{eq-Aug-AA-E-Approx2}
\end{equation}%
This is reasonable, of course, only if \ $\mathring{s}_{k|k}^{\hat{o}_{\hat{L%
}_{k|k}}}(x,l)\cong \hat{s}_{l|\hat{L}_{k|k}}^{k|k}(x)$---i.e., only if the
term \ $\mathring{s}_{k|k}^{\hat{o}_{\hat{L}_{k|k}}}(x,l)$ \ dominates the
summation in \ $\hat{s}_{l|\hat{L}_{k|k}}(x)$ \ for each \ $l\in \hat{L}%
_{k|k}$.

\section{Application to Closely-Spaced Targets \label{A-Application}}

In the GLMB filter, the distribution of birth-targets at time \ $t_{k}$ \ is
assumed to be approximable as an LMB p.d.f., as per Eq. (\ref{eq-Birth}): 
\begin{equation}
\ \mathring{b}_{k|k-1}(\mathring{X})=\delta _{|\mathring{X}|,|\mathring{X}_{%
\mathbb{L}}|}\,\omega _{L_{k|k-1}^{B}}(\mathring{X}_{\mathbb{L}})\,(%
\mathring{s}_{k|k-1}^{B})^{\mathring{X}}.  \label{eq-Apriori}
\end{equation}%
This assumption is implicitly based on a more subtle assumption: The birth
targets are all sufficiently well-separated. \ 

This is not necessarily the case, however. \ In most multitarget tracking
scenarios, the statistically-correlated targets will usually consist of a
small number of well-separated target-clusters, each of which consists of a
small number of closely-spaced targets.\footnote{%
Ground-vehicle convoys typically violate this assumption. \ They can be
addressed using other techniques---e.g., graph-based tracking.} \ 

In recent fast GLMB\ filter implementations such as those reported in \cite%
{Beard2020}, \cite{ShimVo-TSP2023}, multitarget populations are first
partitioned into statistically independent clusters. \ This means that any
cluster of closely-separated (and therefore statistically-correlated)
targets can be processed individually.\ \ (As was pointed out earlier in
Footnote 1, note that if a cluster has at most two targets then it can be
optimally processed using the \textquotedblleft dyadic
filter\textquotedblright\ of \cite{Mahler-IETDyad-2022}.)

In such cases Eq. (\ref{eq-Apriori}) can, for each such newly-appearing
cluster, be replaced by a SLC-LMB p.d.f.: 
\begin{eqnarray}
\ \mathring{b}_{k|k-1}(\mathring{X}) &=&\delta _{|\mathring{X}|,|\mathring{X}%
_{\mathbb{L}}|}\,\omega _{L_{k|k-1}^{B}}(\mathring{X}_{\mathbb{L}%
})\sum_{i\in \mathbb{I}_{k|k-1}}\alpha _{i,k|k-1}^{\mathring{X}_{\mathbb{L}}}
\label{eq-Initial-0} \\
&&\cdot \prod_{l\in L_{k|k-1}^{B}}\mathring{s}_{l,k|k-1}^{i,B}(\mathring{X}%
_{l})  \notag \\
&=&\delta _{|\mathring{X}|,|\mathring{X}_{\mathbb{L}}|}\,\sum_{i\in \mathbb{I%
}_{k|k-1}}\omega _{L_{k|k-1}^{B}}^{i}(\mathring{X}_{\mathbb{L}})\,(\mathring{%
s}_{k|k-1}^{i,B})^{\mathring{X}}  \label{eq-Initial-SLC}
\end{eqnarray}%
where Eq. (\ref{eq-Initial-SLC}) is the GLMB form of Eq. (\ref{eq-Initial-0}%
) with \ 
\begin{equation}
\omega _{L_{k|k-1}^{B}}^{i}(L)=\omega _{L_{k|k-1}^{B}}(L)\,\alpha
_{i,k|k-1}^{L}.\ 
\end{equation}%
Here, \ a) $\mathbb{I}_{k|k-1}\overset{_{\text{abbr.}}}{=}\mathbb{I}%
_{L_{k|k-1}^{B}}$; b) \ $|\mathbb{I}_{k|k-1}|$ \ is small; c) \ $\omega
_{L_{k|k-1}^{B}}(\mathring{X}_{\mathbb{L}})$ \ is the LMB weight defined in
Eq. (\ref{eq-LMB-2}) and \ $L_{k|k-1}^{B}$ \ is the label-set for the
hypothesized birth targets;\ d) the \ $\mathring{s}_{l}^{i,B}(x)$ for \ $%
i\in \mathbb{I}_{k|k-1}$\ \ are hypothesized spatial distributions for
target \ $l$; e) $\sum_{i\in \mathbb{I}_{k|k-1}}\alpha _{i}^{L_{k|k-1}}=1$;
and f) \ $\mathring{X}_{l}$ \ was defined in Eq. (\ref{eq-Big}).\ \ 

The LMB factor \ $\omega _{L_{k|k-1}^{B}}(\mathring{X}_{\mathbb{L}})$ \
allows the hypothesized targets to have non-unity probabilities of
existence. \ (If all probabilities of existence are unity then Eq. (\ref%
{eq-SpecialLMB}) applies.)

In practice, the following procedure can be used for each such cluster:

\begin{enumerate}
\item Reset time the \ $t_{k}$ \ to time \ $t_{1}$, which resets \ $%
\mathring{b}_{k|k-1}(\mathring{X})$ \ to \ $\mathring{b}_{1|0}(\mathring{X})$
\ and \ $\mathring{f}_{k|k-1}(\mathring{X})$ \ to \ $\mathring{f}_{1|0}(%
\mathring{X})=\mathring{b}_{1|0}(\mathring{X})$.

\item Provided that\ \ $|\mathbb{I}_{k|k-1}|$ \ is not too small, additional
target-births are unnecessary in subsequent time-steps, in which case Eqs. (%
\ref{eq-SLC-P-w}-\ref{eq-SLC-P-ss}) can be replaced by, for \ $J\subseteq
L_{k-1|k-1}$, \ 
\begin{eqnarray}
\omega _{k|k-1}(J) &=&\sum_{L\subseteq L_{k-1|k-1}}\omega _{k-1|k-1}(L)
\label{eq-App-P-w} \\
&&\cdot \sum_{o\in \mathbb{O}_{k|k-1}}\omega _{k|k-1}^{S,o}(J|L)\,\alpha
_{o,k-1|k-1}^{L}  \notag \\
\alpha _{o,k|k-1}^{L} &=&\alpha _{o,k|k-1}^{L^{-}}=\sum_{L\subseteq
L_{k-1|k-1}}\rho _{k|k-1}^{S,o}(L^{-}|L)\,\alpha _{o,k-1|k-1}^{L}
\label{eq-App-P-a} \\
\mathring{s}_{k|k-1}^{o}(x,l) &=&\mathbf{1}_{L_{k|k-1}}(l)\,\mathring{s}%
_{k|k-1}^{o,S}(x,l)+\mathbf{1}_{L_{k|k-1}^{B}}(l)\,\mathring{s}%
_{k|k-1}^{B}(x,l)  \label{eq-App-P-s}
\end{eqnarray}
\end{enumerate}

In more detail:

\begin{enumerate}
\item $k=0$: \ $\mathring{f}_{0|0}(\emptyset )=1$ (the scenario is initially
target-free).

\item $k\geq 2$: \ $L_{k|k-1}^{B}=\emptyset $ \ \ (no target births for \ $%
k\geq 2$).

\item $k=1$: \ The target-birth distribution at time \ $t_{1}$ \ is given
by:\ 
\begin{eqnarray}
\mathring{f}_{1|0}(\mathring{X}) &=&\mathring{b}_{1|0}(\mathring{X}) \\
&=&\delta _{|\mathring{X}|,|\mathring{X}_{\mathbb{L}}|}\,\omega
_{L_{1|0}^{B}}(\mathring{X}_{\mathbb{L}})\sum_{i\in \mathbb{I}_{1|0}}\alpha
_{i}^{\mathring{X}_{\mathbb{L}}}\,(\mathring{s}_{l,1|0}^{i,B})^{\mathring{X}}
\\
&=&\delta _{|\mathring{X}|,|\mathring{X}_{\mathbb{L}}|}\,\sum_{i\in \mathbb{I%
}_{1|0}}\omega _{1|0}^{i}(\mathring{X}_{\mathbb{L}})(\mathring{s}%
_{l,1|0}^{i,B})^{\mathring{X}}  \label{eq-YY}
\end{eqnarray}%
where \ $\omega _{1|0}^{i}(L)=\omega _{L_{1|0}^{B}}(\mathring{X}_{\mathbb{L}%
})\,\alpha _{i}^{L}.$
\end{enumerate}

As is clear from Eq. (\ref{eq-YY}), this procedure is just a special case of
a conventional GLMB filter, in which a) the effective initial distribution \ 
$\mathring{f}_{1|0}(\mathring{X})=\mathring{b}_{1|0}(\mathring{X})$ \ is
GLMB\ rather than LMB and b) no births occur in any of the subsequent
time-updates. \ 

What is different (and novel) about it is the fact that the targets are, at
the outset, presumed to be intrinsically correlated, using a SLC model of
correlation in the sense described in Section \ref{a-SLC-AA-Interp}. \ 

By way of contrast and as noted earlier in Remark \ref{Rem-Context}, in the
conventional GLMB\ filter SLC models occur only as consequences of the
measurement collections.

\section{Mathematical Derivations \label{A-Der}}

\subsection{Proof of Eq. (\protect\ref{eq-GLMB-P-Big}) \label{A-Der-AA-1}}

First, as was shown in the derivation of \ Eq. (\ref{eq-GLMB-P}), in Eqs.
(15.282,15.283,15.290) of \cite{Mah-Newbook}, the spatial p.d.f \ $\mathring{%
s}_{k|k-1}^{o,S}(x,l)$ \ initially has the form \ 
\begin{eqnarray}
\hat{s}_{\mathring{X}^{-}}^{o,k|k-1}(x,l) &=&\frac{\sum_{x:(x,l^{\prime
})\in \mathring{X}^{-}}\delta _{l,l^{\prime }}\,\mathring{s}_{l^{\prime
},k-1|k-1}^{o}[\mathring{p}_{S}\mathring{M}_{x}]}{\mathring{s}_{l^{\prime
},k-1|k-1}^{o}[\mathring{p}_{S}]}  \label{eq-Summ} \\
&=&\frac{\sum_{x:(x,l)\in \mathring{X}^{-}}\mathring{s}_{l,k-1|k-1}^{o}[%
\mathring{p}_{S}\mathring{M}_{x}]}{\mathring{s}_{l,k-1|k-1}^{o}[\mathring{p}%
_{S}]}.  \label{eq-Summ2}
\end{eqnarray}%
The summation in the numerator in Eq. (\ref{eq-Summ2}) actually has only a
single term since \ $\mathring{X}^{-}$ is an LFS and so there is only a
single \ $x=\mathring{X}_{l}^{-}\in \mathbb{X}_{0}$ \ such that $(x,l)\in 
\mathring{X}^{-}$ \ where \ $\mathring{X}_{\ast l}^{-}$ \ and \ $\mathring{X}%
_{l}^{-}$ \ are as in Eq. (\ref{eq-Big}).\footnote{%
That is, the numerator is an alternative but inelegant notation for \ $%
\mathring{s}_{l,k-1|k-1}^{o}[\mathring{p}_{S}\mathring{M}_{\mathring{X}%
_{l}^{-}}]$..} \ Thus

\begin{equation}
\hat{s}_{\mathring{X}^{-}}^{o,k|k-1}(\mathring{X}_{l}^{-},l)=\mathring{s}%
_{k|k-1}^{o,S}(\mathring{X}_{\ast l}^{-})=\mathring{s}_{l,k|k-1}^{o,S}(%
\mathring{X}_{l}^{-}),
\end{equation}%
which results in%
\begin{equation}
\mathring{s}_{k|k-1}^{o,S}(x,l)\overset{_{\text{def.}}}{=}\frac{\mathring{s}%
_{l,k-1|k-1}^{o}[\mathring{p}_{S}\mathring{M}_{x}]}{\mathring{s}%
_{l,k-1|k-1}^{o}[\mathring{p}_{S}]}
\end{equation}%
and thus%
\begin{equation}
\mathring{f}_{k|k-1}(\mathring{X})=\delta _{|\mathring{X}|,|\mathring{X}_{%
\mathbb{L}}|}\,\omega _{k|k-1}^{B}(\mathring{X}_{\mathbb{L}}^{+})\sum_{o\in 
\mathbb{O}_{k|k-1}}\tilde{\omega}_{k|k-1}^{o}(\mathring{X}_{\mathbb{L}%
}^{-})\,(\mathring{s}_{k|k-1}^{B})^{\mathring{X}^{+}}(\mathring{s}%
_{k|k-1}^{o,S})^{\mathring{X}^{-}}.
\end{equation}

\subsection{Proof of Eq. (\protect\ref{eq-Normal}) \label{A-Der-AA-2}}

First, from Eq. (\ref{eq-LMB}) we have 
\begin{equation}
\tilde{\omega}_{k|k-1}^{o}(J)=\sum_{L\subseteq \mathbb{L}}\omega
_{k-1|k-1}^{o}(L)\,\omega _{L}(J)
\end{equation}%
where%
\begin{equation}
\omega _{L}(J)=\left( \prod_{l^{\prime }\in L-J}\mathring{s}_{l^{\prime
},k-1|k-1}^{o}[\mathring{p}_{S}^{c}]\right) \left( \prod_{l^{\prime }\in J}%
\mathbf{1}_{L}(l^{\prime })\,\mathring{s}_{l^{\prime },k-1|k-1}^{o}[%
\mathring{p}_{S}]\right)
\end{equation}%
and so%
\begin{eqnarray}
\sum_{J\in \mathbb{L}}\tilde{\omega}_{k|k-1}^{o}(J) &=&\sum_{L\subseteq 
\mathbb{L}}\omega _{k-1|k-1}^{o}(L)\sum_{J\in \mathbb{L}}\omega _{L}(J)
\label{eq-VV} \\
&=&\sum_{L\subseteq \mathbb{L}}\omega _{k-1|k-1}^{o}(L)=1.
\end{eqnarray}%
The identity \ $\sum_{J\in \mathbb{L}}\omega _{L}(J)=1$ \ employed in Eq. (%
\ref{eq-VV}) follows by applying the generalized binomial theorem \cite[Eq.
(3.6)]{Mah-Newbook}, \ \ 
\begin{equation}
(h_{1}+h_{2})^{X}=\sum_{W\subseteq X}h_{1}^{W}\,h_{2}^{X-W},
\end{equation}%
to%
\begin{equation}
1=\sum_{l\in J}(1-a_{l}+a_{l})=\sum_{L\subseteq J}\left( \prod_{l\in
L}(1-a_{l})\right) \left( \prod_{l\in J-L}a_{l}\right)
\end{equation}%
where a) \ $X$ \ has been replaced by \ $J$; b) \ $x\in X$ \ has been
replaced by \ $l\in J\,$; and c) $h_{1}(x)$ \ has been replaced by \ \ $a_{l}%
\overset{_{\text{abbr.}}}{=}\mathring{s}_{l,k-1|k-1}^{o}[\mathring{p}_{S}]$.

\subsection{Proof of Eq. (\protect\ref{eq-FacCov-2}) \label{A-Der-AA-2a}}

Note from Eq. (\ref{eq-Example}) and the product rule for first-order
functional derivatives that the first functional derivative of \ $\mathring{G%
}_{\{l_{1},l_{2}\}}[\mathring{h}]$ \ with respect to \ $\mathring{x}%
_{1}^{\prime }=(x_{1}^{\prime },l_{1}^{\prime })$ \ is 
\begin{equation}
\frac{\delta \mathring{G}_{\{l_{1},l_{2}\}}}{\delta (x_{1}^{\prime
},l_{1}^{\prime })}[\mathring{h}]=\sum_{o}\alpha
_{o}^{\{l_{1},l_{2}\}}\left( \delta _{l_{1}^{\prime },l_{1}}\mathring{s}%
_{l_{1}}^{o}(x_{1}^{\prime })\,\mathring{s}_{l_{2}}^{o}[\mathring{h}]+%
\mathring{s}_{l_{1}}^{o}[\mathring{h}]\,\delta _{l_{1}^{\prime },l_{2}}%
\mathring{s}_{l_{2}}^{o}(x_{1}^{\prime })\right)
\end{equation}%
and so 
\begin{eqnarray}
&&\frac{\delta \mathring{G}_{\{l_{1},l_{2}\}}}{\delta (x_{1}^{\prime
},l_{1}^{\prime })}[1] \\
&=&\sum_{o}\alpha _{o}^{\{l_{1},l_{2}\}}\left( \delta _{l_{1}^{\prime
},l_{1}}\mathring{s}_{l_{1}^{\prime }}^{o}(x_{1}^{\prime })+\delta
_{l_{1}^{\prime },l_{2}}\mathring{s}_{l_{1}^{\prime }}^{o}(x_{1}^{\prime
})\right)  \notag \\
&=&\sum_{o}\alpha _{o}^{\{l_{1},l_{2}\}}\,\mathbf{1}_{\{l_{1},l_{2}%
\}}(l_{1}^{\prime })\,\mathring{s}_{l_{1}^{\prime }}^{o}(x_{1}^{\prime })=%
\mathbf{1}_{\{l_{1},l_{2}\}}(l_{1}^{\prime })\sum_{o}\alpha
_{o}^{\{l_{1},l_{2}\}}\,\mathring{s}_{l_{1}^{\prime }}^{o}(x_{1}^{\prime }).
\end{eqnarray}%
Similarly, 
\begin{equation}
\frac{\delta \mathring{G}_{\{l_{1},l_{2}\}}}{\delta (x_{1}^{\prime
},l_{1}^{\prime })}[1]=\mathbf{1}_{\{l_{1},l_{2}\}}(l_{2}^{\prime
})\sum_{o}\alpha _{o}^{\{l_{1},l_{2}\}}\,\mathring{s}_{l_{2}^{\prime
}}^{o}(x_{2}^{\prime }).
\end{equation}%
The second functional derivative of \ $\mathring{G}_{\{l_{1},l_{2}\}}[%
\mathring{h}]$ with respect to \ $\mathring{x}_{2}^{\prime }=(x_{2}^{\prime
},l_{2}^{\prime })$ \ is \ 
\begin{eqnarray}
&&\frac{\delta ^{2}\mathring{G}_{\{l_{1},l_{2}\}}}{\delta (x_{2}^{\prime
},l_{2}^{\prime })\delta (x_{1}^{\prime },l_{1}^{\prime })}[\mathring{h}] \\
&=&\sum_{o}\alpha _{o}^{\{l_{1},l_{2}\}}\left( \delta _{l_{1}^{\prime
},l_{1}}\mathring{s}_{l_{1}}^{o}(x_{1}^{\prime })\,\delta _{l_{2}^{\prime
},l_{2}}\mathring{s}_{l_{2}}^{o}(x_{2}^{\prime })+\delta _{l_{2}^{\prime
},l_{1}}\mathring{s}_{l_{1}}^{o}(x_{2}^{\prime })\,\delta _{l_{1}^{\prime
},l_{2}}\mathring{s}_{l_{2}}^{o}(x_{1}^{\prime })\right)  \notag \\
&=&\sum_{o}\alpha _{o}^{\{l_{1},l_{2}\}}\left( \delta _{l_{1}^{\prime
},l_{1}}\mathring{s}_{l_{1}^{\prime }}^{o}(x_{1}^{\prime })\,\delta
_{l_{2}^{\prime },l_{2}}\mathring{s}_{l_{2}^{\prime }}^{o}(x_{2}^{\prime
})+\delta _{l_{2}^{\prime },l_{1}}\mathring{s}_{l_{2}^{\prime
}}^{o}(x_{2}^{\prime })\,\delta _{l_{1}^{\prime },l_{2}}\mathring{s}%
_{l_{1}^{\prime }}^{o}(x_{1}^{\prime })\right)
\end{eqnarray}%
\begin{eqnarray}
&=&\sum_{o}\alpha _{o}^{\{l_{1},l_{2}\}}\left( \delta _{l_{1}^{\prime
},l_{1}}\delta _{l_{2}^{\prime },l_{2}}\mathring{s}_{l_{1}^{\prime
}}^{o}(x_{1}^{\prime })\,\mathring{s}_{l_{2}^{\prime }}^{o}(x_{2}^{\prime
})+\delta _{l_{2}^{\prime },l_{1}}\delta _{l_{1}^{\prime },l_{2}}\mathring{s}%
_{l_{2}^{\prime }}^{o}(x_{2}^{\prime })\,\mathring{s}_{l_{1}^{\prime
}}^{o}(x_{1}^{\prime })\right) \\
&=&\sum_{o}\alpha _{o}^{\{l_{1},l_{2}\}}\left( \delta _{l_{1}^{\prime
},l_{1}}\delta _{l_{2}^{\prime },l_{2}}+\delta _{l_{2}^{\prime
},l_{1}}\delta _{l_{1}^{\prime },l_{2}}\right) \,\mathring{s}_{l_{1}^{\prime
}}^{o}(x_{1}^{\prime })\,\mathring{s}_{l_{2}^{\prime }}^{o}(x_{2}^{\prime })
\end{eqnarray}%
\begin{eqnarray}
&=&\sum_{o}\alpha _{o}^{\{l_{1},l_{2}\}}\,\delta
_{\{l_{1},l_{2}\},\{l_{1}^{\prime },l_{2}^{\prime }\}}\,\mathring{s}%
_{l_{1}^{\prime }}^{o}(x_{1}^{\prime })\,\mathring{s}_{l_{2}^{\prime
}}^{o}(x_{2}^{\prime }) \\
&=&\delta _{\{l_{1},l_{2}\},\{l_{1}^{\prime },l_{2}^{\prime
}\}}\sum_{o}\alpha _{o}^{\{l_{1},l_{2}\}}\,\mathring{s}_{l_{1}^{\prime
}}^{o}(x_{1}^{\prime })\,\mathring{s}_{l_{2}^{\prime }}^{o}(x_{2}^{\prime })
\\
&=&\delta _{\{l_{1},l_{2}\},\{l_{1}^{\prime },l_{2}^{\prime
}\}}\sum_{o}\alpha _{o}^{\{l_{1}^{\prime },l_{2}^{\prime }\}}\,\mathring{s}%
_{l_{1}^{\prime }}^{o}(x_{1}^{\prime })\,\mathring{s}_{l_{2}^{\prime
}}^{o}(x_{2}^{\prime }).
\end{eqnarray}%
Finally,\ \ noting that \ 
\begin{equation}
\mathbf{1}_{\{l_{1},l_{2}\}}(l_{1}^{\prime })\,\mathbf{1}_{\{l_{1},l_{2}%
\}}(l_{2}^{\prime })=\delta _{\{l_{1},l_{2}\},\{l_{1}^{\prime
},l_{2}^{\prime }\}},
\end{equation}%
from Eq. (\ref{A-Math-AA-FacCov}) the factorial covariance is, \ 
\begin{eqnarray}
&&\mathring{c}_{\{l_{1},l_{2}\}}^{[2]}(\mathring{x}_{1}^{\prime },\mathring{x%
}_{2}^{\prime }) \\
&=&\delta _{\{l_{1},l_{2}\},\{l_{1}^{\prime },l_{2}^{\prime
}\}}\sum_{o}\alpha _{o}^{l_{1},l_{2}}\,\mathring{s}_{l_{1}^{\prime
}}^{o}(x_{1}^{\prime })\,\mathring{s}_{l_{2}^{\prime }}^{o}(x_{2}^{\prime })
\notag \\
&&-\left( \mathbf{1}_{\{l_{1},l_{2}\}}(l_{1}^{\prime })\sum_{o}\alpha
_{o}^{l_{1},l_{2}}\,\,\mathring{s}_{l_{1}^{\prime }}^{o}(x_{1}^{\prime
})\right) \left( \mathbf{1}_{\{l_{1},l_{2}\}}(l_{2}^{\prime })\sum_{o}\alpha
_{o}^{l_{1},l_{2}}\,\,\mathring{s}_{l_{2}^{\prime }}^{o}(x_{2}^{\prime
})\right)  \notag
\end{eqnarray}%
\begin{equation}
=\delta _{\{l_{1},l_{2}\},\{l_{1}^{\prime },l_{2}^{\prime }\}}\left( 
\begin{array}{c}
\sum_{o}\alpha _{o}^{l_{1},l_{2}}\,\mathring{s}_{l_{1}^{\prime
}}^{o}(x_{1}^{\prime })\,\mathring{s}_{l_{2}^{\prime }}^{o}(x_{2}^{\prime })
\\ 
-\left( \sum_{o}\alpha _{o}^{l_{1},l_{2}}\,\,\mathring{s}_{l_{1}^{\prime
}}^{o}(x_{1}^{\prime })\right) \left( \sum_{o}\alpha _{o}^{l_{1},l_{2}}\,\,%
\mathring{s}_{l_{2}^{\prime }}^{o}(x_{2}^{\prime })\right)%
\end{array}%
\right) .
\end{equation}

\subsection{Proof of Eqs. (\protect\ref{eq-SLC-P-w},\protect\ref{eq-SLC-P-aa}%
) \label{A-Der-AA-3}}

To demonstrate Eq. (\ref{eq-SLC-P-w}), first note that the time-updated GLMB
p.d.f. written in SLC-GLMB form is 
\begin{eqnarray}
\mathring{f}_{k|k-1}(\mathring{X}) &=&\delta _{|\mathring{X}|,|\mathring{X}_{%
\mathbb{L}}|}\,\sum_{o\in \mathbb{O}_{k|k-1}}\omega _{k|k-1}^{o}(\mathring{X}%
_{\mathbb{L}})\,(\mathring{s}_{k|k-1})^{\mathring{X}} \\
&=&\delta _{|\mathring{X}|,|\mathring{X}_{\mathbb{L}}|}\,\omega _{k|k-1}(%
\mathring{X}_{\mathbb{L}})\sum_{o\in \mathbb{O}_{k|k-1}}\alpha _{o,k|k-1}^{%
\mathring{X}_{\mathbb{L}}}\,(\mathring{s}_{k|k-1}^{o})^{\mathring{X}}
\end{eqnarray}%
where%
\begin{eqnarray}
\omega _{k|k-1}(L) &=&\sum_{o\in \mathbb{O}_{k|k-1}}\omega _{k|k-1}^{o}(L) \\
\alpha _{o,k|k-1}^{L} &=&\frac{\omega _{k|k-1}^{o}(L)}{\omega _{k|k-1}(L)}.
\end{eqnarray}%
However, by Eq. (\ref{eq-GLMB-P-w}),%
\begin{equation}
\omega _{k|k-1}^{o}(L)=\omega _{k|k-1}^{B}(L^{+})\,\tilde{\omega}%
_{k|k-1}^{o}(L^{-})
\end{equation}%
and by Eq. (\ref{Eq. (278)}), 
\begin{eqnarray}
\tilde{\omega}_{k|k-1}^{o}(L^{-}) &=&\sum_{L\subseteq L_{k-1|k-1}}\omega
_{k|k-1}^{S,o}(L^{-}|L)\,\omega _{k-1|k-1}^{o}(L) \\
&=&\sum_{L\subseteq L_{k-1|k-1}}\omega _{k|k-1}^{S,o}(L^{-}|L)\,\omega
_{k-1|k-1}(L)\,\alpha _{o,k-1|k-1}^{L}
\end{eqnarray}%
and so Eq. (\ref{eq-SLC-P-w}) follows from%
\begin{eqnarray}
\tilde{\omega}_{k|k-1}(L^{-}) &=&\sum_{o\in \mathbb{O}_{k|k-1}}\tilde{\omega}%
_{k|k-1}^{o}(L^{-}) \\
&=&\sum_{L\subseteq L_{k-1|k-1}}\omega _{k-1|k-1}(L) \\
&&\cdot \sum_{o\in \mathbb{O}_{k|k-1}}\omega _{k|k-1}^{S,o}(L^{-}|L)\,\alpha
_{o,k-1|k-1}^{L}.  \notag
\end{eqnarray}

On the other hand, Eq. (\ref{eq-SLC-P-aa}) follows from 
\begin{eqnarray}
&&\alpha _{o,k|k-1}^{L^{-}} \\
&=&\frac{\sum_{L\subseteq \mathbb{L}}\omega _{k|k-1}^{S,o}(L^{-}|L)\,\omega
_{k-1|k-1}(L)\,\alpha _{o,k-1|k-1}^{L}}{\sum_{L\subseteq L_{k-1|k-1}}\omega
_{k-1|k-1}(L)\sum_{o\in \mathbb{O}_{k|k-1}}\omega
_{k|k-1}^{S,o}(L^{-}|L)\,\alpha _{o,k-1|k-1}^{L}}  \notag
\end{eqnarray}%
\begin{eqnarray}
&=&\sum_{L\subseteq L_{k-1|k-1}}\frac{\omega _{k|k-1}^{S,o}(L^{-}|L)\,\omega
_{k-1|k-1}(L)}{\left( 
\begin{array}{c}
\sum_{L\subseteq L_{k-1|k-1}}\omega _{k-1|k-1}(L) \\ 
\cdot \sum_{o\in \mathbb{O}_{k|k-1}}\,\omega _{k|k-1}^{S,o}(L^{-}|L)\,\alpha
_{o,k-1|k-1}^{L}%
\end{array}%
\right) }\;\;\;\, \\
&&\cdot \alpha _{o,k-1|k-1}^{L}  \notag \\
&=&\sum_{L\subseteq L_{k-1|k-1}}\sigma _{k|k-1}^{o}(L^{-}|L)\,\alpha
_{o,k-1|k-1}^{L}.
\end{eqnarray}

\subsection{Proof of Eqs. (\protect\ref{eq-SLC-B-w},\protect\ref{eq-SLC-B-aa}%
) \label{A-Der--AA-4}}

From Eq. (\ref{eq-GLMB-Meas-Init}) the measurement-updated SLC-GLMB p.d.f.
is, when rewritten in SLC-GLMB\ form, 
\begin{eqnarray}
\mathring{f}_{k|k}(\mathring{X}|Z) &=&\delta _{|\mathring{X}|,|\mathring{X}_{%
\mathbb{L}}|}\sum_{(o,\mathring{\theta})\in \mathbb{O}_{k|k}}\omega
_{k|k}^{(o,\mathring{\theta})}(\mathring{X}_{\mathbb{L}})\,(\mathring{s}%
_{k|k}^{(o,\mathring{\theta})})^{\mathring{X}} \\
&=&\delta _{|\mathring{X}|,|\mathring{X}_{\mathbb{L}}|}\,\omega _{k|k}(%
\mathring{X}_{\mathbb{L}})\sum_{(o,\mathring{\theta})\in \mathbb{O}%
_{k|k}}\alpha _{(o,\mathring{\theta})}^{\mathring{X}_{\mathbb{L}}}\,(%
\mathring{s}_{k|k}^{(o,\mathring{\theta})})^{\mathring{X}}
\end{eqnarray}%
where%
\begin{eqnarray}
\omega _{k|k}(L) &=&\sum_{(o,\mathring{\theta})\in \mathbb{O}_{k|k}}\omega
_{k|k}^{(o,\mathring{\theta})}(L) \\
\alpha _{(o,\mathring{\theta}),k|k}^{L} &=&\frac{\omega _{k|k}^{(o,\mathring{%
\theta})}(L)}{\omega _{k|k}(L)}.
\end{eqnarray}%
From Eq. (\ref{eq-GLMB-Meas-w}), 
\begin{eqnarray}
&&\omega _{k|k}^{(o,\mathring{\theta})}(L) \\
&=&\frac{\omega _{k|k-1}^{o}(L)\,\mathring{\lambda}_{k}^{\mathring{\theta}%
}(L)\prod_{l\in L}\mathring{s}_{l,k|k-1}^{o}[\mathring{L}_{Z_{k}}^{\mathring{%
\theta}}]}{\sum_{L\subseteq \mathbb{L}}\sum_{\mathring{\theta}\in \mathcal{T}%
_{Z_{k}}}\sum_{o\in \mathbb{O}_{k|k-1}}\omega _{k|k-1}^{o}(L)\,\mathring{%
\lambda}_{k}^{\mathring{\theta}}(L)\,\mathring{s}_{l,k|k-1}^{o}[\mathring{L}%
_{Z_{k}}^{\mathring{\theta}}]}  \notag
\end{eqnarray}%
\begin{eqnarray}
&=&\frac{\omega _{k|k-1}(L)\,\alpha _{o,k|k-1}^{L}\,\mathring{\lambda}_{k}^{%
\mathring{\theta}}(L)\prod_{l\in L}\mathring{s}_{l,k|k-1}^{o}[\mathring{L}%
_{Z_{k}}^{\mathring{\theta}}]}{\left( 
\begin{array}{c}
\sum_{L\subseteq \mathbb{L}}\omega _{k|k-1}(L)\,\sum_{o\in \mathbb{O}%
_{k|k-1}}\alpha _{o,k|k-1}^{L}\, \\ 
\cdot \sum_{\mathring{\theta}\in \mathcal{T}_{Z_{k}}}\mathring{\lambda}_{k}^{%
\mathring{\theta}}(L)\,\omega _{k|k-1}^{o}(L)\prod_{l\in L}\mathring{s}%
_{l,k|k-1}^{o}[\mathring{L}_{Z_{k}}^{\mathring{\theta}}]%
\end{array}%
\right) } \\
&=&\omega _{k|k-1}(L)\,\alpha _{o,k|k-1}^{L}\,\rho _{k|k-1}^{(o,\mathring{%
\theta})}(Z_{k}|L)
\end{eqnarray}%
and thus Eq. (\ref{eq-SLC-B-w}) follows this and from%
\begin{eqnarray}
\omega _{k|k}(L) &=&\sum_{o\in \mathbb{O}_{k|k-1}}\,\sum_{\mathring{\theta}%
\in \mathcal{T}_{Z_{k}}}\omega _{k|k}^{(o,\mathring{\theta})}(L) \\
&=&\omega _{k|k-1}(L)\sum_{o\in \mathbb{O}_{k|k-1}}\sum_{\mathring{\theta}%
\in \mathcal{T}_{Z_{k}}}\alpha _{o,k|k-1}^{L}\,\rho _{k|k-1}^{(o,\mathring{%
\theta})}(Z_{k}|L).
\end{eqnarray}

On the other hand, Eq. (\ref{eq-SLC-B-aa}) follows from 
\begin{equation*}
\alpha _{(o,\mathring{\theta}),k|k}^{L}=\frac{\omega _{k|k}^{(o,\mathring{%
\theta})}(L)}{\sum_{o\in \mathbb{O}_{k|k-1}}\sum_{\mathring{\theta}\in 
\mathcal{T}_{Z_{k}}}\omega _{k|k}^{(o,\mathring{\theta})}(L)}
\end{equation*}%
\begin{eqnarray}
&=&\frac{\omega _{k|k-1}(L)\,\alpha _{o,k|k-1}^{L}\,\rho _{k|k-1}^{(o,%
\mathring{\theta})}(Z_{k}|L)}{\omega _{k|k-1}(L)\,\sum_{o\in \mathbb{O}%
_{k|k-1}}\alpha _{o,k|k-1}^{L}\sum_{\mathring{\theta}\in \mathcal{T}%
_{Z_{k}}}\rho _{k|k-1}^{(o,\mathring{\theta})}(Z_{k}|L)} \\
&=&\frac{\alpha _{o,k|k-1}^{L}\,\rho _{k|k-1}^{(o,\mathring{\theta}%
)}(Z_{k}|L)}{\sum_{o\in \mathbb{O}_{k|k-1}}\alpha _{o,k|k-1}^{L}\sum_{%
\mathring{\theta}\in \mathcal{T}_{Z_{k}}}\rho _{k|k-1}^{(o,\mathring{\theta}%
)}(Z_{k}|L)}.
\end{eqnarray}

\section{Conclusions \label{A-Concl}}

In this paper it has been demonstrated that:

\begin{enumerate}
\item GLMB p.d.f.'s can be interpreted as straightforward generalizations of
LMB p.d.f.'s to statistically correlated target populations, based on an
implicit presumption of \textquotedblleft simple labeled
correlation\textquotedblright\ (SLC) models of multitarget correlation. \ 

\item SLC\ models represent the kinematic uncertainty of individual targets
using random spatial p.d.f.'s instead of the usual deterministic spatial
p.d.f.'s.

\item SLC\ models generalize GLMB multitarget state models to account for 
\textit{a priori} multitarget correlation---i.e., correlations not
attributable to collected measurements.

\item The GLMB\ filter can be reformulated as a SLC-GLMB filter.

\item SLC\ models appear to be most appropriate for addressing
target-clusters consisting of small numbers of closely-spaced targets.
\end{enumerate}

\end{document}